\documentclass{IEEEtran}
\usepackage{pgfplots}
\pgfplotsset{compat=1.18}
\usepackage{pgfplotstable}
\usepgfplotslibrary{groupplots}
\usepackage[]{xcolor}
\usepackage{colortbl}
\usepackage{tikz}
\usepackage{cite}
\usepackage{svg}
\usepackage{stfloats}
\usepackage[font=footnotesize]{caption}
\usepackage[font=footnotesize]{subcaption}
\usepackage{multirow}
\usepackage{amssymb}
\usepackage{booktabs}
\usepackage{amsmath,amssymb,amsfonts}
\usepackage{algpseudocode}
\usepackage[ruled,vlined, linesnumbered]{algorithm2e}
\usepackage{optidef}
\usepackage{graphicx}

\usepackage{graphicx,color}
\usepackage{dirtytalk}
\usepackage{textcomp}
\usepackage{setspace}
\usepackage{ragged2e}
\usepackage{bbm}
 \AtBeginEnvironment{bmatrix}{\setlength{\arraycolsep}{3pt}}
\newtheorem{remark}{Remark}

\begin{document}

\newcommand{\roa}[1]{\textcolor{blue}{[Ramoni: #1]}}

\title{Goal-Oriented Interference Coordination in 6G In-Factory Subnetworks}
\author{Daniel Abode, \IEEEmembership{Member, IEEE}, Pedro Maia de Sant Ana, \IEEEmembership{Member, IEEE}, Ramoni Adeogun, \IEEEmembership{Senior Member, IEEE}, Alexander Artemenko, \IEEEmembership{Member, IEEE}, Gilberto Berardinelli, \IEEEmembership{Senior Member, IEEE}
\thanks{The work by Daniel Abode was supported by the Horizon 2020 research and innovation programme under the Marie Skłodowska-Curie grant agreement No. 956670. The work by Pedro Maia de Sant Ana, Ramoni Adeogun, and Gilberto Berardinelli, was supported by the HORIZON-JU-SNS-2022-STREAM-B-01-03 6G-SHINE project (grant agreement No. 101095738). Ramoni Adeogun is also supported by the HORIZON-JU-SNS-2022-STREAM-B-01-02 CENTRIC project (grant agreement No. 101096379)}
\thanks{Daniel Abode is with the Department of Electronic Systems, Aalborg University, Aalborg 9220 Denmark (e-mail: danieloa@es.aau.dk). }
\thanks{Pedro Maia de Sant Ana is with the Corporate Research and Advanced Engineering, Robert Bosch GmbH, Renningen 71272, Germany (e-mail: Pedro.MaiadeSantAna@de.bosch.com). }
\thanks{Ramoni Adeogun is with the Department of Electronic Systems, Aalborg University, Aalborg 9220 Denmark (e-mail: ra@es.aau.dk). }
\thanks{Alexander Artemenko is with the Corporate Research and Advanced Engineering, Robert Bosch GmbH, Renningen 71272, Germany (e-mail: alexander.artemenko@de.bosch.com).}
\thanks{Gilberto Berardinelli is with the Department of Electronic Systems, Aalborg University, Aalborg 9220 Denmark (e-mail: gb@es.aau.dk).}}

\markboth{Journal of \LaTeX\ Class Files,~Vol.~14, No.~8, August~2021}%
{Shell \MakeLowercase{\textit{et al.}}: A Sample Article Using IEEEtran.cls for IEEE Journals}


\maketitle

\begin{abstract}
Subnetworks are expected to enhance wireless pervasiveness for critical applications such as wireless control of plants, however, they are interference-limited due to their extreme density. This paper proposes a goal-oriented joint power and multiple sub-bands allocation policy for interference coordination in 6G in-factory subnetworks. Current methods for interference coordination in subnetworks only focus on optimizing communication metrics, such as the block error rate, without considering the goal of the controlled plants. This oversight often leads to inefficient allocation of the limited radio resources. To address this, we devise a novel decentralized inter-subnetwork interference coordination policy optimized using a Bayesian framework to ensure the long-term stability of the subnetwork-controlled plants. Our results show that the proposed decentralized method can support more than twice the density of subnetwork-controlled plants compared to centralized schemes that aim to minimize the block error rate while reducing execution complexity significantly.

\end{abstract}


\section{Introduction}
\label{sec:introduction}

\IEEEPARstart{T}{he} vision of Industry 4.0 and beyond involves smart factories enabled by cognitive, autonomous, flexible, and reconfigurable production entities. Connectivity is crucial for realizing this vision, as the production entities are equipped with various sensors, controllers, and actuators that share information for their functionalities \cite{chen_smart_2018}. The inflexibility and high cost of deploying wired connections have increased the interest in adopting wireless technology for communication between the sensors, controllers and actuators in production systems. To take advantage of the market opportunity, the wireless community continues to push boundaries for improved reliability, capacity, and reduced latency \cite{seferagic_survey_2020}. 5G radio technology has introduced support for high data rates and mission-critical services such as ultra-reliable low latency communication (URLLC) for controlling vehicles or industrial machinery. However, it is already apparent that 5G has limitations in meeting the stringent latency and reliability requirements of near real-time industrial automation at the field level such as motion control \cite{bartolin-arnau_private_2023}. 

The limitations of 5G motivate research toward the next wireless generation, 6G, expected to expand wireless capabilities to support critical services \cite{VH2020}. The 6G \say{network of network} vision \cite{VH2020} integrates subnetworks at the edge to enhance wireless pervasiveness, providing extremely reliable local wireless service inside entities like vehicles, humans, robots and the like \cite{EU2021,Adeogun2020,Gilberto2023}. These subnetworks will offload local communication needs from the central 6G/5G network, while the central network may manage the subnetworks within its coverage area for improved spectral efficiency. From initial studies, high mutual interference between subnetworks has been identified as a major limitation to the capacity of subnetworks deployment \cite{Gilberto2023,Adeogun2020}. Previous solutions to managing interference in cellular systems may not be applicable because of the extreme density, mobility and large scale of subnetworks. For instance, in the case of in-factory subnetworks, many industrial modules carrying subnetworks may be close together on the factory floor. To enable subnetworks towards 6G, developing novel and effective methods to manage inter-subnetwork interference is crucial. In \cite{Abode2024PC,Du2023,Adeogun2020x,Li2023}, heuristic and data-driven methods for sub-band allocation and power control have been studied to mitigate inter-subnetwork interference. In sub-band allocation, the available bandwidth is divided into a finite number of sub-bands, which are then optimally reused among the subnetworks. The power control problem involves optimizing the transmit power of the transmitters in the subnetworks to maximize the network's spectral efficiency. 

Existing works on interference coordination for in-factory subnetworks focus on optimizing wireless communication metrics such as spectral efficiency and block error rate (BLER) \cite{Abode2024PC, Du2023, Adeogun2020x, Li2023}. However, such approaches often overlook the specific characteristics of the underlying control system, whose goal is to maintain the stability of the plant. Recent studies show that control systems may require highly reliable wireless communication in some state conditions and tolerate looser communication requirements in others, while still achieving their goal \cite{Saeed2023, Pedro2022, LIMA20202634, Chamaken2010, Wang2021, Onur2024}. 
For example, a wirelessly controlled mobile robot navigating a complex trajectory may require a high-speed and highly reliable control loop to maintain stability. In contrast, during simpler tasks, such as traversing a straight, clutter-free path, the communication system can operate under looser reliability requirements. These variations underscore the need for a co-design approach that aligns the wireless communication objectives with the specific goals of the control system. In interference-limited environments, such as dense subnetworks, ignoring this interaction can lead to over-provisioning and misallocation of the available radio resources. For instance, pursuing overly stringent wireless performance metrics can result in unnecessary interference, degrading the performance of neighboring subnetworks.

This paper takes a novel approach to tackling inter-subnetwork interference in subnetwork-controlled plants by exploring the interplay between the dynamics of these systems and the constraints of limited radio resources. We propose a novel decentralized transmit power and sub-band allocation protocol inspired by recent advances in goal-oriented Radio Resource Management (RRM). Our findings demonstrate the significant practical benefits of a co-design framework in managing inter-subnetwork interference within dense subnetwork-controlled plants compared to traditional control-agnostic methods.

\begin{figure}[]
    \centering
    \includegraphics[height=6.8cm]{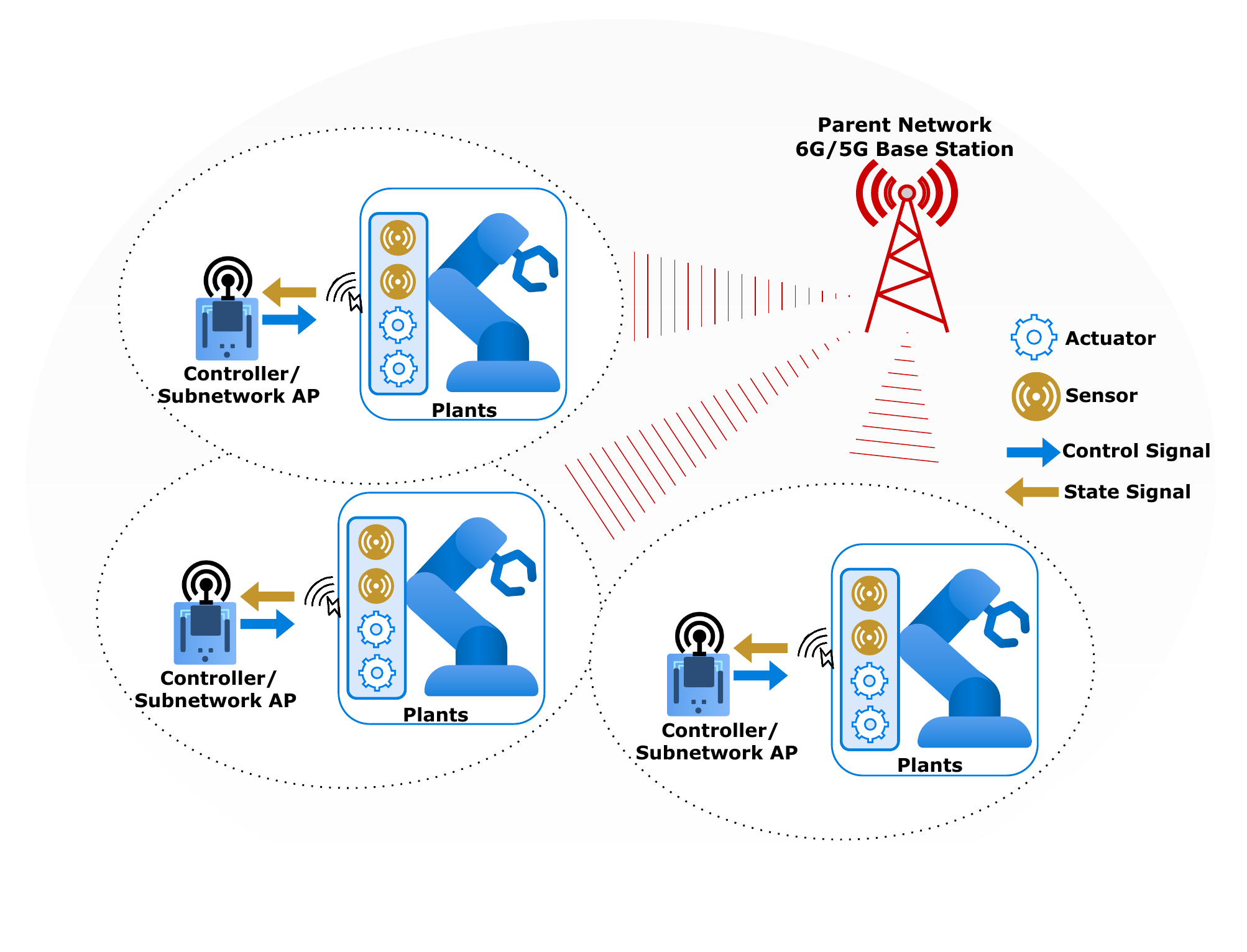}
    \caption{In-factory subnetworks supporting closed-loop control of their associated plants at the field level. The parent network is responsible for centralized radio resource allocation. However, the subnetworks may autonomously manage interference in a decentralized configuration.}
    \label{fig:inFs2}
\end{figure}

\subsection{Review of Related Works}

\subsubsection{Wireless Network Control Systems} 
Traditionally, wireless network control systems (WNCS) have been implemented with ad-hoc technologies such as WiFi \cite{Zand2012} and more recently cellular technologies, prominently 5G private network \cite{bartolin-arnau_private_2023}. A major challenge is how the WiFi access point (AP), or the 5G private network may schedule transmission and serve the large number of devices in the WNCS. The current scheduling protocols utilized in WiFi and 5G are limited for WNCS due to the decoupling of wireless network scheduler design from the dynamic objective of control systems \cite{Pedro2022}. Recent works have shed light on the inefficiencies in the current standardized scheduling protocols and have proposed goal-oriented scheduling algorithms that consider the dynamics of the controlled system. The authors in \cite{Pedro2022} examined a multi-user setting of multiple plants served by a single base station. They propose a joint network and control scheduler to allocate frequency resources to plants to maximize overall control system stability. Their proposed method supported twice the number of plants compared to traditional algorithms such as round robin with the same radio resources. Likewise, the authors in \cite{Eisen2020} proposed a control-optimal wireless resource scheduling algorithm and demonstrated its robust scalability to an increasing number of devices compared to an optimal packet error rate scheduler. The authors in \cite{Wang2021} established a linear relationship between the stability of the controlled system and the average age of state information. This was utilized to optimize the inter-arrival rate of state information and code length reducing the joint control and communication cost. The study by \cite{LIMA20202634} explored power control for multiple plants organized in a multicellular architecture. They introduced a reinforcement learning algorithm that significantly reduced control costs compared to methods optimizing for radio metrics. Collectively, the results from \cite{Saeed2023, Pedro2022, LIMA20202634, Chamaken2010, Wang2021}  and other related studies underscore the potential of integrating the dynamics of controlled systems into RRM for WNCS.  

\subsubsection{Interference Mitigation in 6G In-X Subnetworks}
The RRM challenges in subnetworks-enabled WNCS differ from the architecture mentioned above, where the main RRM challenge involves scheduling multiple plants and devices to a central factory network. In 6G subnetwork control systems illustrated in Figure~\ref{fig:inFs2}, where a subnetwork consists of wireless sensors, actuators and an AP colocated with the controller, two interference problems abound. One is intra-subnetwork interference resolved by scheduling the sensors and actuators in the subnetwork on orthogonal radio resource slots similar to the problem in \cite{Pedro2022, Eisen2020}. The most crucial is the inter-subnetwork interference as multiple subnetworks operate simultaneously on shared radio resources.

The primary focus for inter-subnetwork interference mitigation has been to develop sub-band allocation and transmit power control solutions with lower execution costs. Execution costs include the time and frequency resources needed to obtain channel information, and the algorithm computation complexity. The interference management algorithm must be quick to adapt to the continuously changing channel conditions due to the mobility and varying density of subnetworks. This fast adaptability is crucial as subnetworks are expected to facilitate deterministic and sub-millisecond latency communication within entities  \cite{Gilberto2023}. The authors in \cite{Adeogun2020x} proposed three distributed methods for dynamically selecting channels in subnetworks including the nearest neighbour conflict avoidance (NNCA). In NNCA, subnetworks sense the channels occupied by $K-1$ neighbours and select an available unoccupied channel. They showed that NNCA performs similarly to the centralized graph colouring benchmark, achieving better spectral efficiency than the other distributed methods, albeit with higher sensing and signalling costs. The problem of joint power and sub-band allocation was studied using reinforcement learning in \cite{Du2023, Adeogun}. In \cite{Du2023}, the authors proposed a multi-agent reinforcement learning scheme to dynamically adapt the selection of transmit power levels and sub-bands based on measured received signal strength indicators on the channels. The authors in \cite{Adeogun} proposed a multi-agent Q learning algorithm which realized better performance and was more robust to sensing interval and switching delay compared to the greedy algorithm. In \cite{Mandelli2021}, the authors proposed a link adaptation scheme to reduce the transmit power used in subnetworks while transferring the unused time-frequency resources in non-congested cells to congested cells.  The authors in \cite{GB2023} proposed a hybrid approach, such that the interference management may be centralized or decentralized depending on the quality of the backhaul link to the parent network, the density of the subnetworks, and the local sensing capacity of the subnetworks. In \cite{Li2023}, the authors proposed a centralized sequential iterative sub-band allocation (SISA) scheme that minimizes the sum of interference to signal ratio and demonstrated its effective performance in achieving a lower BLER than the decentralized benchmark algorithm. It has been shown that centralized interference management algorithms in which the parent 6G/5G network manages the inter-subnetwork interference are more efficient, however, they incur more cost of signalling between the subnetworks and the parent network \cite{Li2023, Adeogun2022}. 

It is important to note the following limitations of current interference mitigation techniques in 6G in-X subnetworks which motivates this study. 
\begin{itemize}
    \item While centralized approaches, such as SISA, offer significant improvements in maximizing the reliability of subnetworks compared to decentralized schemes, they come with significant execution costs. These methods rely on information about the mutual inter-subnetwork interference channel, which is difficult to obtain in practice considering the constantly changing channel due to subnetwork mobility and traffic.
    \item In the sub-band allocation problem, it is of particular interest that subnetworks requiring more radio resources may utilize more than one sub-band. However, prior studies are generally limited to single sub-band allocation.
    \item Although most studies refer to in-factory use cases of subnetworks for wireless control of plants, they do not consider the dynamics of the underlying controlled plants. Meanwhile, as previously stated, considering the dynamic response of the underlying applications has been identified as a promising approach to efficiently managing radio resources \cite{Saeed2023, Pedro2022, LIMA20202634, Chamaken2010, Wang2021}. 
\end{itemize}  

In response to the limitations of previous research on mitigating inter-subnetwork interference, particularly for subnetwork controlled systems, our study in \cite{Daniel2024} conducted an initial investigation into control-aware transmit power control for in-factory subnetworks. We proposed the control-aware channel-independent (CICA) transmit power control that eliminates the need for costly sensing and signalling of radio information. We show that CICA performs better than control-agnostic methods that require accurate channel information for power control decisions to maximize spectral efficiency. However, power control alone may not be sufficient for effectively managing interference between subnetworks, especially in densely deployed subnetworks that support plants with strict communication requirements. Hence, this study proposes a general problem formulation and a novel decentralized solution for joint transmit power and multiple sub-bands allocation to effectively manage interference in the dense deployment of the subnetworks for control systems.

\subsection{Major Contributions}
Our major contributions are as follows:
\begin{enumerate}
    \item We formulate the inter-subnetwork interference coordination problem as a joint power and multiple sub-bands allocation scheme that minimizes the finite horizon control costs for the subnetwork-controlled plants, without the need for cumbersome mutual inter-subnetwork interference channel measurement.
    \item We propose a decentralized inter-subnetwork interference coordination solution that consists of a logistic function for transmit power allocation and a piece-wise constant function for multi sub-bands allocation that can be easily implemented with low computation complexity. We further propose a Bayesian optimization framework using a multi-objective tree-structured parzen estimator to optimize the trainable parameters of the logistic function and the piece-wise constant function. 
    \item We conduct extensive simulation experiments to justify the efficient performance and scalability of the proposed solution compared to traditional centralized benchmark algorithms that are designed to minimize the BLER.
\end{enumerate}

\subsection{Structure of the paper}
The rest of the paper is organized as follows. We present some preliminary knowledge in Section \ref{section:preliminaries}. Section \ref{section_systemmodel} details the model of the subnetwork-controlled system. We discuss our problem formulation and solution in Section \ref{section_problem}. The simulation experiment assumption and results discussion are detailed in Sections \ref{section_settings} and \ref{section_result} respectively. We noted the limitations of the current study and proposed directions for future study in section \ref{sect_futurework}. Finally, we conclude in Section \ref{section_conclusion} and discuss some considerations for future works.

\section{Preliminaries}
\label{section:preliminaries}
\subsection{Control System}
A control plant can be modelled with a state space equation, with rate of change of state $\Dot{\mathbf{x}} = \mathbf{A}\mathbf{x} + \mathbf{B}\mathbf{u} + \mathbf{w}$ \cite{Pedro2022,Michael2003-rg}, where $\mathbf{x} \in \mathbb{R}^{q}$ is the state space vector with $q$ variables. $\mathbf{u} = -\mathbf{k}(\mathbf{x} - \mathbf{x}_{\text{desired}}) \ \in \mathbb{R}^{r}$ is the linear feedback control vector of $r$ variables with a gain of $\mathbf{k} \in \mathbb{R}^{q \times r}$ and $\mathbf{x}_{\text{desired}}$ is a vector of desired state. $\mathbf{w} \in \mathbb{R}^{q}$ is the system disturbance modelled as Gaussian noise with covariance $\mathbf{\Sigma} \in  \mathbb{R}^{q \times q}$ and mean of zero. $\mathbf{A} \in \mathbb{R}^{q\times q}$ is the system state matrix that dictates how the state vector of the system changes from time $t$ to $t+1$ when $\mathbf{u} = \mathbf{0}$. $\mathbf{B} \in \mathbb{R}^{q\times r}$ indicates how the state of the system changes from $t$ to $t+1$ when control action $\mathbf{u}$ is applied. We assume that even with no action, $\mathbf{x} \to \infty \text{ as } t \to \infty$, i.e. $\mathbf{A}$ is unstable.
We assume $\mathbf{x}_{\text{desired}} = \mathbf{0}$, resulting in the system dynamics
\begin{equation}
    \Dot{\mathbf{x}} = (\mathbf{A} - \mathbf{B}\mathbf{k})\mathbf{x} + \mathbf{w}
    \label{eqn:system_dynamics}
\end{equation}
Note that in a discrete form $\Dot{\mathbf{x}} = (\mathbf{x}^{(t+dt)} - \mathbf{x}^{(t)})/dt$, where $dt$ is the sampling interval.
The feedback control gains $\mathbf{k}$ is generally optimally predesigned \cite{Pedro2022,An2021}, for example, using a well-known technique such as the Linear Quadratic Regulator \cite{Michael2003-rg}. The linear quadratic regulator is defined to find the optimal control policy that minimizes a weighted quadratic cost of the plant states and the control action over an infinite horizon. The weighted quadratic cost $\chi^{(t)}$ at time $t$, referred to as the instantaneous LQR cost, is expressed as
\begin{equation}
    \chi^{(t)} =\mathbf{x}^{(t)^*} \mathbf{Q} \mathbf{x}^{(t)} + \mathbf{u}^{(t)^*} \mathbf{R} \mathbf{u}^{(t)}
    \label{eqn:inst_lqrcost}
\end{equation}
where \say{$*$} denotes the transpose. And
\begin{equation}
    \Bar{\chi} = \int_0^{\infty} \chi^{(t)} dt,
    \label{eqn:lqrcost}
\end{equation}
is the infinite horizon LQR cost. We refer interested readers to \cite{Michael2003-rg} for a detailed explanation of the Linear Quadratic Regulator. The LQR cost is a measure of the stability of the plant, i.e. the higher the LQR cost the less stable the plant.

We define a measure of the control cost over a finite horizon as the average of the instantaneous LQR cost over a finite horizon $T$ as \cite{Pedro2022} 
\begin{equation}
    \Tilde{\chi} = \frac{1}{T}\sum_{t=0}^{T-1} \chi^{(t)}.
    \label{eqn:finiteLQR}
\end{equation}
In this paper, we call $\Tilde{\chi}$, the finite horizon control cost and $\chi$, the instantaneous control cost. The goal of a subnetwork-controlled plant is to minimise this control cost, which captures how effectively the system maintains its states within acceptable bounds and minimises energy or actuation effort, as in the two components of \eqref{eqn:inst_lqrcost} \cite{Pedro2022}. The control cost can sufficiently describe the performance of a goal-oriented subnetwork-controlled plant because it directly quantifies how well the system achieves its control objectives within the constraints of the wireless communication network.

\begin{remark}
Some works consider the co-design of the control law $\mathbf{u}$ and wireless resource allocation, generally referred to as joint communication and control co-design such as in \cite{Chamaken2010, Saeed2023, lima_model-free_2022}. However, the scope of this study is to manage the interference in large-scale deployment of subnetworks to achieve the minimum operation cost given a pre-designed control law, similar to the works in \cite{LIMA20202634, An2021, Pedro2022}.
\end{remark}

\subsection{Communication Model}
To model the communication error in this study, we adopted the BLER model from the finite-blocklength information theory in  \cite{Petar2018} popularly used to model error probability in URLLC designed to support ultra-reliable transmission of small packets within a latency constraint. 
Given the signal-to-interference-plus-noise ratio (SINR), \(\gamma_k\), of a channel block \(k\) with bandwidth \(B_k\), and a transmit latency constraint of \(\tau\), the number of channel uses, \(\Pi\), is defined as the number of instances a communication channel is utilized to transmit information. Specifically, each symbol transmission constitutes one channel use, and for a bandwidth-limited system, the number of channel uses is given by \(\Pi = 2B_k \tau\) \cite{Petar2018}. For the transmission of a packet of \(b = D + D'\) bits, where \(D\) is the payload and \(D'\) is the metadata, the block error probability for a \(1/2\) code rate is approximated by \cite{Petar2018}:
\begin{equation}
    \epsilon_k(\Pi,\gamma_k,b) = Q\left( \frac{\frac{\Pi}{2}\log_2(1 + \gamma_k) - b + \frac{1}{2}\log_2 \Pi}{\sqrt{\Pi V(\gamma_k)}}\right),
    \label{eqn:errormodel}
\end{equation}
where 
\begin{equation}
    V(\gamma_k) = \frac{\gamma_k(\gamma_k+2)}{2(\gamma_k+1)^2} \log^2_2 e 
\end{equation}
is the channel dispersion. If we consider joint encoding of the data and metadata, the probability of successful reception is $(1 - \epsilon_k(\Pi,\gamma_k,b))$. 

The successful reception indicator of data on channel block $k$ in sub-band $l$ denoted as $\mathbbm{P}_{k,l}$ is then
\begin{equation}
    \mathbbm{P}_{k,l} = 
    \begin{cases}
        True, & \text{if} \ \Bar{\epsilon} < 1 - \epsilon_k(\Pi,\gamma,b) \\
        False, & \text{otherwise},
    \end{cases}
    \label{eqn:errorindicatorblock}
\end{equation}
where $\Bar{\epsilon} \sim \mathcal{U}(0,1)$ is sampled from a standard uniform distribution. 
In subnetworks, the available bandwidth is divided into $K$ channel blocks while the channels are grouped in $L$ sub-bands. Each subnetwork can operate on $\Bar{L} \leq L$ sub-bands, dividing the payload into $\Bar{L} \times K$ bits. To complete the transmission error model, we define a boolean variable $\mathbbm{P}$ for the successful reception of data transmitted over the operating sub-bands as
     \begin{equation}
     \mathbbm{P} = 
    \begin{cases}
        True, \ \text{if} \ \mathbbm{P}_{k,l} \ \forall \ k = \{1,2,\ldots,K\}, \ l = \{1,2,\ldots,\Bar{L}\}\\
        False, \ \text{otherwise},
    \end{cases}
    \label{eqn:errorindicatorsubbands}
\end{equation}
which emphasizes that the transmitted data is successfully received if all the data transmitted over the channel blocks in the sub-bands are successfully delivered.


\subsection{Bayesian Optimization using Multi-objective Tree Structured Parzen Estimator}
\label{section:BO}
Bayesian optimization (BO) is a machine learning approach used to solve optimization problems of the form \cite{frazier_tutorial_nodate}
\begin{mini}|l|[1]
{{\scriptsize \begin{aligned} \mathbf{y} \in \mathcal{A} \end{aligned}}}{\mathbf{f(y)}}
{\label{eqn:bayetut}}{}
\end{mini}
sequentially, where the set $\mathcal{A}$ is defined as the feasible region for the optimization problem in \eqref{eqn:bayetut}, encompassing all possible values of $\mathbf{y}$ that satisfy the constraints of the system. BO is often called sequential model-based optimization (SMBO) because of its sequential approach to solving optimization problem. It is particularly useful when the function $\mathbf{f(y)}$ is expensive to evaluate, especially in cases where the gradient and structure are unavailable. Also, the function may be nonlinear and non-convex. In essence, $\mathbf{f(y)}$ is treated as a black box \cite{frazier_tutorial_nodate}.  It has been suggested that SMBO has specific properties that make it a practical machine-learning optimization algorithm for RRM \cite{maggi_bayesian_2021}. It offers rapid convergence and safer exploration compared to reinforcement learning. In addition, it naturally allows embedding prior expert knowledge in the optimisation process \cite{maggi_bayesian_2021}. A detailed tutorial on Bayesian optimization for RRM, with a case study of open loop power control configuration, is presented in \cite{maggi_bayesian_2021}.

The following steps summarise the Bayesian optimization of $\mathbf{f(y)}$. First, an initial set of $S$ hyperparameter configurations \(\mathbf{y}_1, \mathbf{y}_2, \ldots, \mathbf{y}_S\) is randomly generated to evaluate $\mathbf{f(y)}$. This results in a dataset of $\{\mathbf{y}_1, \mathbf{f(y_1)}), (\mathbf{y}_2, \mathbf{f(y_2)}), \ldots, (\mathbf{y}_S, \mathbf{f(y_S)})\}$. The next step involves fitting a probabilistic model (surrogate model) to the observed data. The most commonly used probabilistic models in the literature are Gaussian Processes (GP), Random Forests (RF), and Tree-Parzen Estimators (TPE) \cite{Bergstra2011}. Subsequently, an acquisition function determines the next set of configurations to be evaluated. In a single objective problem, the common acquisition function includes Expected Improvement, Probability of Improvement, or Confidence Bound \cite{frazier_tutorial_nodate}. However, in multiple objective problems, it is common to use the Expected Hyper-Volume Improvement (EHVI) \cite{Ozaki2022}. The next step is the evaluation process, in which the objective is assessed with the new observations. Finally, in the update step, the dataset is updated with the new data and the probabilistic model is updated accordingly. The steps from acquisition to update are repeated until a stopping criterion, such as the maximum number of evaluations or convergence condition is met.

 In this work, the optimization variables involve sub-band selection, which is discrete, and transmit power selection, which can be continuous. Therefore, we consider using Tree-Parzen estimators, which are more suitable when the optimization variables must be sampled from a combination of continuous and discrete spaces \cite{pmlr-v28-bergstra13}. Also, we consider multiple objectives, since different sets of parameters may be needed depending on whether we are interested in improving the average performance, the worst-case performance, or a tradeoff between both. In multi-objective problem $\mathbf{f(y)} = \{\mathbf{f_1(y)}, \mathbf{f_2(y)}, \ldots, \mathbf{f_V(y)}\}$, where $V$ is the number of objectives. In the following, we discuss the Multi-objective Tree-Parzen Estimator (MOTPE) BO method employed in this study. 
 
 As described in Algorithm \ref{algMOTPE}, MOTPE takes a set of $\mathcal{S}$ observations, $\{(\mathbf{y}_1,\mathbf{f(y_1)}), (\mathbf{y_2},\mathbf{f(y_2)}), \ldots, (\mathbf{y}_{\mathcal{S}},\mathbf{f}(\mathbf{y_{\mathcal{S}}})\}$. Then these observations are split into good observations and bad observations. In practice, a specific quantile value $\gamma$ is chosen and the observations are sorted and split given the chosen quantile \cite{Bergstra2011}.  Two probability density functions $l(\mathbf{y})$ and $g(\mathbf{y})$ are used to model $p(\mathbf{y} \mid \mathbf{f(y)})$ as in \cite{Ozaki2020}

\begin{equation}
    p(\mathbf{y} \mid \mathbf{f(y)}) = 
\begin{cases} 
l(\mathbf{y})& \text{if } \mathbf{y} \in \text{Good observations} \\
g(\mathbf{y}) & \text{if }  \mathbf{y} \in \text{Bad observations},
\end{cases}
\label{eqn:lygy}
\end{equation}
\( l(\mathbf{y})\) and \( g(\mathbf{y}) \) are density functions estimated using Parzen Estimators, also known as kernel density estimator (KDE) \cite{Shuhei2023}.

\begin{algorithm}
\caption{Bayesian optimization using MOTPE}
\label{algMOTPE}
\KwIn{$O = \{(\mathbf{y}^{(i)},\mathbf{f(y^{(i)})}) \mid i = 1, 2, \ldots, \mathcal{S}\}, \mathcal{T}, \mathcal{C}, \gamma$ \tcp* {Observations, Number of iterations, candidates per iteration, quantile}}
\For{$t \gets 1$ \KwTo $\mathcal{T}$}{
    $(O_l, O_g) \gets \texttt{SPLIT\_OBSERVATIONS}(O, \gamma)$\;
    Construct $l(\mathbf{y}), g(\mathbf{y})$ from $O_l, O_g$\;
    \tcp{Sample $\mathcal{C}$ candidates for $\mathbf{y}$}
    $C(\mathbf{y}) \gets \{\mathbf{y}^{(c)} \sim l(\mathbf{y}) \mid c = 1, \ldots, \mathcal{C}\}$\; 
    \tcp{Approximate best candidate}
    $\mathbf{y}' \gets \arg\max_{\mathbf{y} \in C(\mathbf{y})} \frac{l(\mathbf{y})}{g(\mathbf{y})}$\; 
    $O \gets O \cup \{(\mathbf{y}', \mathbf{f(y')})\}$\;
}
\Return{the set of non-dominated solutions in $O$}\;
\end{algorithm}


 In \cite{Ozaki2020}, the authors presented the performance of MOTPE over different standard benchmark Bayesian optimization tasks, and they showed that $0.05 \leq \gamma \leq 0.2$ were reasonable values for most of the considered optimization problems. The complete MOTPE algorithm is presented in Algorithm \ref{algMOTPE}. Generally, in one iteration, the observations are split and the $l(\mathbf{y}),g(\mathbf{y})$ are constructed as in lines $2,3$ in Algorithm \ref{algMOTPE} following (\ref{eqn:lygy}). Some candidate values of $\mathbf{y}$ are sampled from the good observation distribution $l(\mathbf{y})$, then the value of $\mathbf{y}$ that maximizes the expected hyper-volume improvement (EHVI) which can be approximated as $l(\mathbf{y})/g(\mathbf{y})$ (denoted as $\mathbf{y}'$) is selected and added to the observation set. At the end of the iterations, the set of values of $\mathbf{y}$ in the observation set that achieves the best performance for the objectives $\mathbf{f(y)}$, referred to as the set of non-dominated solutions is selected. Since $\mathbf{f(y)}$ are multi-objective, it is possible that $\mathbf{y}$ that minimizes one objective may not minimize the other objective. In any way, selecting the actual value of $\mathbf{y}$ from the non-dominated set used in the system would depend on what objectives or combination of objectives is more important for the system.


\begin{figure}[]
    \centering
    \includegraphics[height=8.7cm]{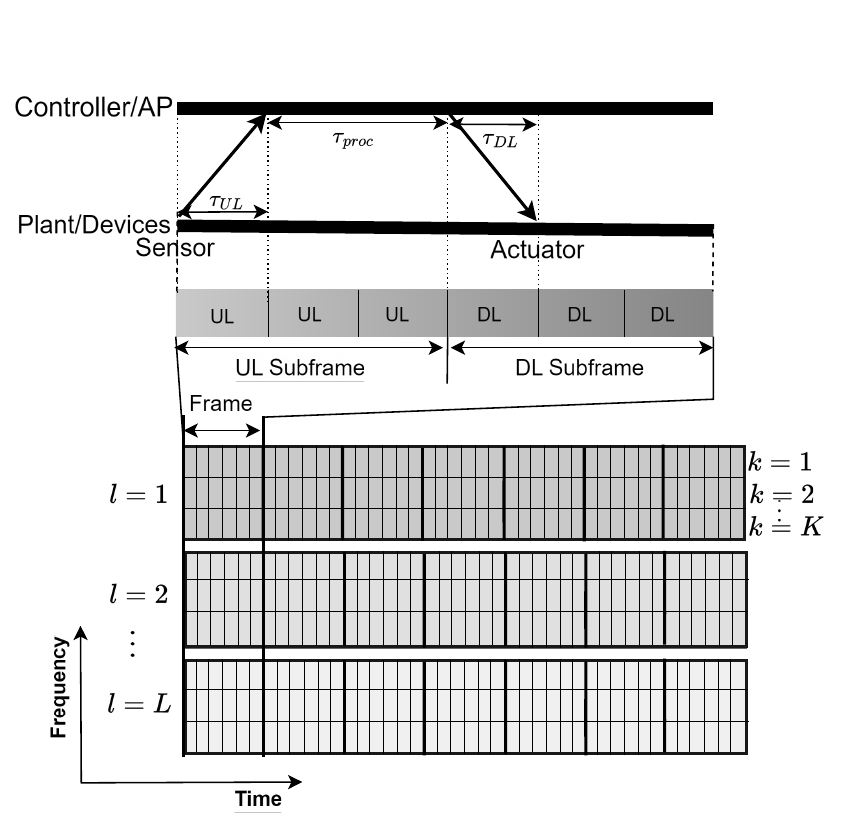}
    \caption{Illustration of the In-factory subnetwork control system model depicting the TDD frame structure for the periodic traffic, and the shared frequency-time resource grid with $L$ sub-bands, where each sub-band consists of $K$ channels.}
    \label{fig:signalling}
\end{figure}

\section{In-Factory Subnetwork Control System Model}
\label{section_systemmodel}
The In-factory subnetwork control system consists of $N$ subnetworks each supporting the closed-loop control of their associated plants as shown in Figure \ref{fig:inFs2}. We assume each subnetwork carries a controller with AP capability supporting one or more plants. Figure \ref{fig:signalling} shows the radio resource grid structure shared by the $N$ subnetworks, the uplink (UL) and downlink (DL) data communication between the sensor, actuator of the plant and the AP/controller in a subnetwork. The available bandwidth is divided into $L$ sub-bands, and one sub-band consists of $K$ channels. The operation time is divided into frames. The UL and DL transmission units are arranged in subframes in a time division duplex (TDD) manner, and the transmit data is divided across the channels in the operating sub-band or sub-bands of the subnetwork. In the literature, the transmission time for UL, $\tau_{UL}$ and DL, $\tau_{DL}$ are generally specified as sub-milliseconds for closed-loop control applications \cite{Adeogun2020, Li2023}. Each plant has sensors that collect observations about the states $\mathbf{x}$ of the plants and transmit them wirelessly UL to the AP/controller in the subnetwork. The controller processes the received plants' state in $\tau_{proc} ~\text{(sec)}$ and wirelessly transmits the control signal DL to the actuator. However, since the subnetwork is operating on limited radio resources shared with many other subnetworks, the wireless channel for both the UL and DL are imperfect due to noise and interference and may be subjected to unsuccessful transmission as deduced from \eqref{eqn:errormodel}. This transmission error probability depends on the SINR ($\gamma_k$), channel use ($\Pi$), and latency constraint ($\tau$). In this paper, we assume that undelivered packets are not re-transmitted. 

\subsection{The SINR Model}
To complete the communication model, we must model the SINR for both the UL and the DL. For simplification, we assume that the sensors are collocated with a single wireless transceiver, and the actuator also has a single wireless transceiver. Hence in the following model of the SINR, the subnetwork consists of two devices, the UL device (sensor) and the DL device (actuator). In addition, we assume that the UL and DL are time-synchronized across all subnetworks, hence UL transmission does not interfere with a DL transmission. The power received by the AP in subnetwork $n \in \{1,2,\ldots,N\}$ due to UL transmission from the sensor with transmit power $p_{n,k,l,\text{UL}}^{(t)} $ over a channel $k$ in sub-band $l$ characterized by complex small-scale fading gain $h_{n,k,l,\text{UL}}^{(t)}$, and large scale fading $\rho_{n,k,l,\text{UL}}^{(t)}$ in time $t$ is given as 
\begin{equation}
    q_{n,k,l,\text{UL}}^{(t)} = p_{n,k,l,\text{UL}}^{(t)}  |h_{n,k,l,\text{UL}}^{(t)}| ^2  \rho_{n,k,l,\text{UL}}^{(t)}.
\end{equation}
The received interference power at subnetwork $n$ due to a sensor in subnetwork $n' \in \mathcal{N}^{(t)}_{n,k,l} \subset \{1,2,\ldots,N\}$ transmitting simultaneously with power $p_{n',k,l,\text{UL}}^{(t)}$ over the same channel is given as 
\begin{equation}
    q_{n',n,k,l,\text{UL}}^{(t)} = p_{n',k,l,\text{UL}}^{(t)}  |h_{n',n,k,l,\text{UL}}^{(t)}| ^2  \rho_{n',n,k,l,\text{UL}}^{(t)}.
\end{equation}
Where $\mathcal{N}_{n,k,l}^{(t)}$ is the set of subnetworks with UL devices operating on channel block $k$ in subband $l$ at the same time $t$ as subnetwork $n$.
We can then model the UL SINR as 
\begin{equation}
    \gamma_{n,k,l,\text{UL}}^{(t)} = \frac{q_{n,k,l,\text{UL}}^{(t)}}{\sum\limits_{n' \in \mathcal{N}_{n,k,l}^{(t)}} q_{n',n,k,l,\text{UL}}^{(t)} + \sigma^2}, 
    \label{eqn:SINR}
\end{equation}
where $\sigma^2$ is the thermal noise power. The DL SINR $\gamma_{n,k,l,\text{DL}}^{(t)}$ of the link between the AP in subnetwork $n$  and the actuator can be modelled similarly. 

\subsection{The Communication-Control Model}

Given the SINR, $q_{n',n,k,l,\text{UL}}^{(t)}$, of the UL communication from the sensor to the controller, the probability of successful reception of the sensor data can be determined using the communication model described in \eqref{eqn:errorindicatorsubbands}. This probability, denoted as $\mathbbm{P}_{\text{UL}}$, quantifies the reliability of UL communication. If the UL transmission succeeds, the controller computes and transmits a new control action to the actuator. The corresponding probability of successful reception of the control data, $\mathbbm{P}_{\text{DL}}$, can also be derived by \eqref{eqn:errorindicatorsubbands} using $\gamma_{n,k,l,\text{DL}}^{(t)}$

However, when either the sensor-to-controller (UL) or controller-to-actuator (DL) communication fails, the actuator operates based on an estimated control action, $\mathbf{\Bar{u}} = -\mathbf{k}\mathbf{\Bar{x}}$. Here, $\mathbf{\Bar{x}}$ represents the last known state of the plant, and $-\mathbf{k}\mathbf{\Bar{x}}$ is assumed to be the most recent successfully transmitted control signal. This fallback mechanism ensures the continuity of the control process, albeit with reduced performance compared to the fully closed-loop operation.
To model this behavior, we define the plant dynamics as a switched system supported by the subnetwork, governed by the following equations:

\begin{equation}
    \mathbf{\Dot{x}} = 
    \begin{cases}
        (\mathbf{A} - \mathbf{B}\mathbf{k})\mathbf{x} + \mathbf{w},\ & \text{closed-loop: if } \mathbbm{P}_{\text{UL}} \cap \mathbbm{P}_{\text{DL}}, \\
        (\mathbf{A} - \mathbf{B}\mathbf{k})\mathbf{\Bar{x}} + \mathbf{w},\ & \text{open-loop: otherwise}.
    \end{cases} 
 \label{eqn:loop}
\end{equation}

Here, $\cap$ denotes the logical AND operation, indicating that the closed-loop operation is achieved only when both UL and DL transmissions succeed. In contrast, the open-loop mode is used when either or both communication links fail. Essentially, in the closed-loop mode, the plant operates with the most recent state feedback $\mathbf{x}$, enabling precise control based on real-time data. In the open-loop mode, the system relies on the estimated state $\mathbf{\Bar{x}}$, which inherently introduces a lag and may lead to performance degradation due to outdated information.

The communication-control model highlights the interplay between communication reliability and control performance. The probabilities $\mathbbm{P}_{\text{UL}}$ and $\mathbbm{P}_{\text{DL}}$ are critical parameters that determine the likelihood of the plant achieving closed-loop operation. As these probabilities decrease due to inter-subnetwork interference, the system increasingly operates in open-loop mode, which may lead to instability or suboptimal performance.

\section{Dynamic Interference Coordination for In-Factory Subnetwork Control Sytems}
\label{section_problem}
In this section, we will describe the problem formulation and our proposed solution for control-aware decentralized interference coordination (CADIC) in the In-factory subnetwork control system. We introduce a decentralized approach, allowing each subnetwork to select radio resources dynamically, depending on the stability condition of the controlled plant, using a centrally trained policy. 

\subsection{Control-Aware Interference Coordination}
We consider a joint transmit power and sub-band selection problem. The subnetwork $n$ must select a transmit power level, $p_n^{(t)}$ as a function $g_p(\cdot)$ of the instantaneous control cost of the plant. In addition, the subnetwork $n$ must select one or more sub-bands denoted with the variable $\mathbf{s}_n^{(t)} \in \mathbbm{B}^L$ using the function $g_{\mathbf{s}}(\cdot)$ of the instantaneous control cost and the received signal strength indicator (RSSI) measured on each subband, i.e. the sensed cumulative interference plus noise on the sub-bands. $\mathbf{s}_n^{(t)}$ is a boolean vector with a length equal to the number of sub-bands, $L$. In the vector, the selected sub-bands are represented as $1$s and the unselected sub-bands as $0$s. For example, if $L=3$, a subnetwork with $\mathbf{s}_n^{(t)} = [1,0,0]$ operates on the first sub-band. Our objective is to select $p_n^{(t)}$ and $\mathbf{s}_n^{(t)}$ that minimizes the finite horizon control cost $\Tilde{\chi}_n$ $\forall n \in \{1,2,\ldots,N\}$. However, defining an objective for each subnetwork would lead to $N$ objectives and will require solving the optimization problem every time $N$ changes, which is not scalable. Hence, we define an arbitrary function to combine the objectives of all the subnetworks, $f\left(\{\Tilde{\chi}_n \ \forall n \in \{1,2,\ldots,N\}\} \right)$. Where $f(\cdot)$ could be a single objective, such as the mean of $\{\Tilde{\chi}_n \ \forall n \in \{1,2,\ldots,N\}\}$, or multiple objectives such as the mean and the maximum of $\{\Tilde{\chi}_n \ \forall n \in \{1,2,\ldots,N\}\}$. The proposed problem formulation is then defined as follows
\begin{mini!}|l|[3]
{{\scriptsize \begin{aligned}p_n^{(t)}, \mathbf{s}_n^{(t)}\end{aligned}}}{f\left(\{\Tilde{\chi}_n \ \forall n \in \{1,2,\ldots,N\}\} \right)}
{\label{eqn:optiobj}}{}
\addConstraint{p_n^{(t)} = g_p(\chi_n^{(t)})}
\label{eqn:congp}
\addConstraint{0  \leq p_n^{(t)} \leq p_{\text{max}}}{}
\label{eqn:conP}
\addConstraint{\mathbf{s}_n^{(t)} = g_{\mathbf{s}}\left(\chi_n^{(t)}, I_{n,1,\text{UL}}^{(t)}, I_{n,2,\text{UL}}^{(t)}, \ldots, I_{n,L,\text{UL}}^{(t)}\right)}{}
\label{eqn:congs}
\addConstraint{ \mathbf{s}_n^{(t)}[l] \in \{0, 1\} \quad \text{for } l = 1, 2, \ldots, L}
\label{eqn:conS1}
\addConstraint{1 \leq \sum_{l=0}^L s_{n,l}^{(t)} \leq L}
\label{eqn:conS2}
\end{mini!}
The objective \eqref{eqn:optiobj} is to optimize the selection of subnetwork transmit power and operating sub-bands to achieve the goal of the subnetwork-controlled plant, which is to minimize the long term control costs.
Constraints \eqref{eqn:congp} and \eqref{eqn:congs} establish that the power allocation decision and the sub-band selection decision are functions of the instantaneous control cost. Constraints \eqref{eqn:conS1} and \eqref{eqn:conS2} imply that the subnetwork may select one or more sub-bands, where the sub-band selected is indicated by $1$ and the unselected sub-band is indicated by $0$ in the sub-band selection vector $\mathbf{s}_{n}$. The transmit power decision is also constrained by a maximum allowable transmit power per subnetwork as in \eqref{eqn:conP}. Note that the transmit power decision is the total per transmission, hence it is equally split over all the operating sub-bands. In addition, as noted in \eqref{eqn:congs}, the sub-band decision also depends on $I_{n,l,\text{UL}}^{(t)}$, the cumulative interference plus noise for all sub-band $l \in \{1, 2, \ldots, L\}$ measured at subnetwork $n$ at time $t$. It is the sum of 
\begin{equation}
 I_{n,k,l,\text{UL}}^{(t)} = \sum\limits_{n' \in \mathcal{N}_{n,k,l}^{(t)}} q_{n',n,k,l,\text{UL}}^{(t)} + \sigma^2,
\end{equation}
the denominator term in \eqref{eqn:SINR} over all the channels $k$ in the sub-band, i.e.
\begin{equation}
    I_{n,l,\text{UL}}^{(t)} = \sum\limits_{k=1}^K I_{n,k,l,\text{UL}}^{(t)}
\end{equation}
Note that the interference is measured only in the UL, hence in the following we generally omit the subscript UL. We focus on the interference measured in the UL phase rather than the DL phase. This is because radio resource optimization is more critical for the UL, which is more prone to communication errors. Essentially, the UL data is typically larger than the DL data, as it carries sensor data which are generally larger than control data in the DL \cite{Pedro2022}. Additionally, only a subset of subnetworks that successfully complete UL communication will need to initiate DL communication, resulting in less interference in the DL.
Likewise, as it is already established that the decisions are per subnetwork and per time unit, we have also omitted the subscript $n$, and $(t)$ for conciseness. Next, we discuss the model for the transmit power function and sub-band selection functions.

\subsubsection{Transmit power function $p = g_p(\chi)$}
First, we consider the power control subproblem, where $p = g_p(\chi)$. Apparently, from \eqref{eqn:SINR}, increasing the transmit power of a subnetwork has the potential to increase the SINR for a constant interference term. Increased SINR has a positive impact on closing the control loop according to (\ref{eqn:errormodel}), hence it is intuitive that $g(\chi)$ should grow as $\chi$ increases. However, $g(\chi)$ must satisfy the transmit power constraint in (\ref{eqn:conP}). Specifically, we desire a function such that $g(\chi) \rightarrow p_{max}$ as $\chi \rightarrow \infty$, and $g(\chi) \rightarrow 0$ as $\chi \rightarrow -\infty$.
Hence, to model $g(\chi)$, we consider a function that grows as $\chi$ increases while having a supremum at $p_{max}$. Examples of functions that exhibit these desired properties are logistic functions with a supremum of $p_{max}$ and piecewise linear functions clipped at $p_{max}$. For instance, the piecewise linear function is used in open loop power control (OLPC) \cite{Ubeda2008} to model the transmit power as a linear function of the path loss clipped at maximum transmit power. However, in this study we consider the parameterized logistic function \cite{ART2005} with a supremum of $p_{max}$ given as  
\begin{equation}
    g_p(\chi; k_0, k_1) = \frac{p_{max}}{1 + \exp(-k_0(\chi -  k_1))}.
    \label{eqn:logistic}
\end{equation}
With a fixed $p_{max}$, the model (\ref{eqn:logistic}) has two learnable parameters $k_0, k_1$ that can be adjusted to set the response of the function to $\chi$. $k_0$ is referred to as the logistic growth rate, which determines the steepness of the function, while $k_1$ is the $\chi$ value where $g_p(\chi; k_0, k_1) = 0.5 \times p_{max}$. Compared to a piecewise linear function with clipping, the logistic function offers high versatility, allowing it to model non-linear relationships and the minimum transmit power more effectively by tuning only two parameters. Moreover, the logistic function can be parameterised to shape like the piecewise linear function but with better smoothness, providing a more refined control.

\subsubsection{Sub-band selection function $g_{\mathbf{s}}(\cdot)$}
Intuitively, a subnetwork should select the sub-band where it experiences the least interference, this would achieve the best SINR according to (\ref{eqn:SINR})  which generally implies a lower block error probability (\ref{eqn:errormodel}).  Likewise, a subnetwork may select more than one sub-band if a single sub-band is insufficient to guarantee lower block error probability to ensure the successful closed-loop control required to drive the plant to a low control cost. When operating on multiple sub-bands,  
the data size per channel is reduced, which has a positive impact on lowering the block error probability  (\ref{eqn:errormodel}). Note we assume the data is split equally over multiple sub-bands and individually encoded per sub-band. To model $g_{\mathbf{s}}(\cdot)$, we consider two functions. First, a function $g_{RL}( I_1^{(t)}, I_{2}^{(t)}, \ldots, I_{L}^{(t)})$ that ranks the sub-bands using the cumulative interference. We define the rank of sub-band $l$ by subnetwork $n$ at time $t$ as its index in the ordered set of the measured interference plus noise on the sub-band averaged over time, i.e.,
    \[
\text{Rank}_{l}^{(t)} = \text{argsort}_l \left\{ \frac{1}{t}\sum_{i=0}^{t-1}I_{1}^{(t)}, \frac{1}{t}\sum_{i=0}^{t-1}I_{2}^{(t)}, \ldots, \frac{1}{t}\sum_{i=0}^{t-1}I_{L}^{(t)} \right\}
\]

Hence, 
\begin{equation}
    g_{RL}( I_1^{(t)}, I_{2}^{(t)}, \ldots, I_{L}^{(t)}) = \{\text{Rank}_{l}^{(t)} \  \forall l \in 1,2, \ldots, L\}
    \label{eqn:rank}
\end{equation}

It is important to note that the sub-band rankings are based on the long-term average of interference measurements for each sub-band. This approach helps avoid the ‘ping-pong’ effect that might arise from using instantaneous interference power.

The second part of $g_{\mathbf{s}}(\cdot)$ is a function that returns the number of sub-bands $\bar{L}$ that are to be used by the subnetwork given $\chi$. We denote this function as $g_{\bar{L}}(\chi)$. Since $g_{\bar{L}}(\chi)$ can only be discrete and $1 \leq g_{\bar{L}}(\chi) \leq L$ as emphasized by  (\ref{eqn:conS1}), (\ref{eqn:conS2}), a suitable function is a parameterized non-negative piece-wise constant function, also called the step function \cite{piecewise}. A parameterized non-negative piecewise constant function is a function that takes constant values over specific intervals that can be parameterized and is always greater than or equal to zero. We can describe $g_{\bar{L}}(\chi)$ as a parameterized non-negative piecewise constant as 

\begin{equation}
    g_{\bar{L}}(\chi; {z_1, z_2, \ldots, z_{L-1}}) = 
    \begin{cases}
        1, \  \text{if} \  0 \leq \chi < z_1, \\
        2, \  \text{if} \  z_1 \leq \chi < z_2, \\  
        \vdots \\
        L, \ \text{if} \  z_{L-1} \leq \chi < \infty. \\
    \end{cases}
    \label{eqn:gL}
\end{equation}
We then have learnable parameters $z_1, z_2, \ldots, z_{L-1}$, which are the pairwise disjoint intervals that determine when the number of sub-bands should increase or decrease. The pairwise disjoint intervals must be carefully learnt, since for example if $z_2$ is set to a $\chi$ value that is too low, many subnetworks may operate on two sub-bands, and the SINR may be diminished due to interference. Now, putting together the components of  $ \mathbf{s} = g_{\mathbf{s}}\left(\chi, I_1^{(t)}, I_{2}^{(t)}, \ldots, I_{L}^{(t)}; {z_1, z_2, \ldots, z_{L-1}}\right)$, we have
\begin{equation}
\mathbf{s}[l]  =
    \begin{cases}
          1 \ \text{for} \ l \ \in  \ g_{RL}( I_1^{(t)}, I_{2}^{(t)}, \ldots, I_{L}^{(t)})[1:\bar{L}], \\
          0, \ \text{for} \ l \ \notin	\ g_{RL}( I_1^{(t)}, I_{2}^{(t)}, \ldots, I_{L}^{(t)})[1:\bar{L}].
    \label{eqn:subband_dec}
    \end{cases}
\end{equation}
Where $g_{RL}( I_1^{(t)}, I_{2}^{(t)}, \ldots, I_{L}^{(t)})[1:\bar{L}]$ is a subset of the first $\bar{L}$ values of the set in \eqref{eqn:rank}. As evident in \eqref{eqn:subband_dec}, the rationale of the proposed sub-band selection function, $g_{\mathbf{s}}(\cdot)$ is that the subnetwork will always use the sub-bands with lower interference based on rank of the sub-bands \eqref{eqn:rank} (i.e., in a greedy-like approach), but the number of sub-bands is selected according to \eqref{eqn:gL}.

Finally, we can now redefine our joint power and sub-band allocation problem. Given our trainable models $p = g_p(\chi; k_0, k_1)$, $\mathbf{s} = g_{\mathbf{s}}\left(\chi, I_{l}^{(t)}, \ldots, I_{L}^{(t)}; {z_1, z_2, \ldots, z_{L-1}}\right)$  with trainable parameters $k_0, k_1$ and $z_1, z_2, \cdots, z_{L-1}$, we complete our problem formulation by rewriting (\ref{eqn:optiobj}) as 
\begin{mini!}|l|[3]
{{\scriptsize \begin{aligned}
    k_0, k_1, z_1, z_2, \cdots, z_{L-1}
\end{aligned}}}{f\left(\{\Tilde{\chi}_n \ \forall n \in \{1,2,\ldots,N\}\}\right),}
{\label{eqn:optiobj2}}{}
\addConstraint{p = g_p(\chi; k_0, k_1)}{}
\label{eqn:conP31}
\addConstraint{\mathbf{s} = g_{\mathbf{s}}\left(\chi, I_1^{(t)}, I_{2}^{(t)}, \ldots, I_{L}^{(t)}; {z_1, z_2, \ldots, z_{L-1}}\right).}{}
\label{eqn:conP32}
\end{mini!}

As earlier noted, the selection of $k_0, k_1$ and $z_1, z_2, \cdots, z_{L-1}$ is non-trivial, as, for example, wrong values may encourage the subnetwork to greedily transmit with maximum power or transmit on all sub-bands even though the associated plant is stable, hence causing high interference and diminishing the overall system performance. Next, in section \ref{section:proposedsolution}, we present a Bayesian optimization approach to tune the parameters $k_0, k_1$ and $z_1, z_2, \cdots, z_{L-1}$ in \eqref{eqn:optiobj2}. 

\begin{remark}
    In this work, we use the estimate of the instantaneous control cost to measure the control plant's stability. However, an estimate of the plant's state was used in some other literature on control-aware RRM such as in \cite{LIMA20202634}. Also, Age of Information (AoI) or a counter from the last successful transmission is used in \cite{Pedro2022, Eisen2020}. In any way, our methodology can generalize to any of this by substituting the instantaneous control cost for any alternative measure that characterizes the stability of the control system.
\end{remark}

\subsection{Proposed Solution using Bayesian Optimization}
\label{section:proposedsolution}
To solve for the parameters in (\ref{eqn:optiobj2}), we consider applying the Bayesian optimization using the MOTPE described in section \ref{section:BO}. The BO method is an effective optimization approach for this problem as it allows the integration of expert knowledge into the process. For instance, we can clearly define the search space for the tunable parameters, establish the dependencies between them, and initialize the optimization variables with promising values. To apply BO, we formulate $f\left(\{\Tilde{\chi}_n \ \forall n \in \{1,2,\ldots,N\}\}\right)$ as two objectives, the mean of $\Tilde{\chi}_n \ \forall \ n$, $f_\mu(\{\Tilde{\chi}_n \  \forall \ n\})$ and the maximum,  $f_{\text{max}}(\{\Tilde{\chi}_n \ \forall \ n\})$.  We also define the set of search spaces $K_0 \subset \mathbb{R}^+$, $K_1  \subset \mathbb{R}^+$, $Z_1  \subset \mathbb{R}^+_{\geq k_1}$, $Z_2  \subset \mathbb{R}^+_{\geq z_1}$, $\ldots$, $Z_{L-1} \subset \mathbb{R}^+_{\geq z_{L-2}}$ for the parameters $k_0$, $k_1$, $z_1$, $z_2$, $\ldots$, $z_{L-1}$ respectively. Such that $k_0 \in K_0 \subset \mathbb{R}^+$, $k_1 \in K_1  \subset \mathbb{R}^+$,  $z_1 \in Z_1  \subset \mathbb{R}^+_{\geq k_1}$, $z_2 \in Z_2  \subset \mathbb{R}^+_{\geq z_1}$, $\ldots$ , $z_{L-1} \in Z_{L-1} \subset \mathbb{R}^+_{\geq z_{L-2}}$. Note the dependency between the search spaces denoted as $\mathbb{R}^+_{\geq *}$, this means, for example, the search space $Z_2  \subset \mathbb{R}^+_{\geq z_1}$  starts from the value of variable $z_1$. TPE has been identified as the preferred variant of BO methods in handling such dependency between search spaces, or tree-structured search spaces \cite{Shuhei2023}, further motivating our choice of the algorithm.
Hence (\ref{eqn:optiobj2}) is reformulated as
\begin{mini}
    {{\scriptsize
      \begin{aligned}
        k_0 &\in K_0, \\
        k_1 &\in K_1, \\
        z_1 &\in Z_1, \\
        z_2 &\in Z_2, \\
        &\vdots \\
        z_{L-1} &\in Z_{L-1}
      \end{aligned}}
      }
    { f_\mu(\{\Tilde{\chi}_n \ \forall \ n\}) \ , f_{\text{max}}(\{\Tilde{\chi}_n \ \forall \ n\}) }
    {\label{eqn:optiobj3}}{}
    \addConstraint{(\ref{eqn:conP31}), (\ref{eqn:conP32}).}
\end{mini}
We consider multiple objectives since we are interested in designing the joint sub-bands and power allocation algorithm to optimize the average performance of all the plants and/or to account for the worst case. Note that we can directly make an analogy between (\ref{eqn:optiobj3}) and (\ref{eqn:bayetut}) such that the array of parameters is mapped as $\mathbf{y} \leftarrow [k_0, k_1$ and $z_1, z_2, \cdots, z_{L-1}]$ and the array of objectives is $\mathbf{f}(\mathbf{y}) \leftarrow [f_\mu(\{\Tilde{\chi}_n \ \forall \ n\}) \ , f_{\text{max}}(\{\Tilde{\chi}_n \ \forall \ n\})]$. 

\begin{algorithm}
    \caption{CADIC Parameter Tuning using MOTPE BO}\label{alg_CADIC}
    \KwIn{$O = \{\}$, $S$, $T$, $E$, $\mathcal{T}, \mathcal{C}$, $\gamma$ \tcp*{Observations, number of startup trials, finite horizon, episodes, number of iterations, number of candidates per iteration, quantile}}
    
    \For{$b \gets 1 \dots \mathcal{S}$}{ 
        \tcp{Randomly pick parameters}
        $\mathbf{y}^{(b)} \gets [k_0, k_1, z_1, z_2, \dots, z_{L-1}]^{(b)}$ \\ 
        $\Tilde{\chi}_n  \gets \{\}$ \tcp*{Initialize a set for the finite horizon control cost}
        
        \For{$e \gets 1 \dots E$}{
            Initialize random deployment and start subnetwork control systems
            
            \For{$t \gets 1 \dots \mathcal{T}$}{
                $\chi_n^{(t)} \ \forall n $;  \tcp{Compute instantaneous control cost using (\ref{eqn:lqrcost})}
                $p_n^{(t)} = g_p(\chi_n^{(t)}; k_0, k_1) \ \forall n$; \tcp{Compute transmit power}
                $\mathbf{s}_n^{(t)} = g_{\mathbf{s}}\left(\chi_n^{(t)}, I_{n,1}^{(t)}, \dots, I_{n,L}^{(t)}; z_1, \dots, z_{L-1}\right) \ \forall n$ \tcp{Compute sub-bands}
                Enable transmission and update plant states using (\ref{eqn:loop})
            }
            $\Tilde{\chi}_n^{(e)} \ \forall n$ \tcp{Compute finite-horizon control cost using (\ref{eqn:finiteLQR})}
        }
        
        $\mathbf{f}(\mathbf{y}) \gets [f_\mu(\{\Tilde{\chi}_n \ \forall n\}) \ , f_{\text{max}}(\{\Tilde{\chi}_n \ \forall n\})]$ \tcp*{Compute objectives}
        $O \gets O \cup (\mathbf{y}^{(b)}, \mathbf{f}(\mathbf{y}^{(b)}))$ \tcp*{Update observation}
    }
    
    \For{$t \gets 1 \dots \mathcal{T}$}{
        Compute Bayesian optimization as in lines 3 to 7 in Algorithm \ref{algMOTPE}, computing objectives using lines 4 to 12 above
    }
    
    \KwOut{$\mathbf{y}' \gets [k_0, k_1, z_1, z_2, \dots, z_{L-1}]'$ \tcp*{That minimizes the objectives}}
\end{algorithm}

\subsubsection{CADIC parameter tuning}
 The optimization procedure to tune the parameters of CADIC is shown in Algorithm \ref{alg_CADIC}. The first step is to collect observations as in lines $2$ to $12$ of the algorithm. In practice, these observations can be gathered centrally from the beginning of an experimental or simulation model of subnetwork-controlled plants. That is, for one observation, values of $k_0, k_1, z_1, z_2, \ldots, z_{L-1}$ which make up $\mathbf{y}$ are randomly selected from their respective search spaces $K_0, K_1, Z_1, Z_2, \ldots, Z_{L-1}$. Based on this value, the transmit power function and sub-band allocation function $g_{\mathbf{s}}(\cdot)$ are determined and used by the subnetworks to run the plants over $E$ episodes and $T$ time steps to collect the finite horizon control cost. The MOTPE algorithm starts with these observations and over several trials $\mathcal{T}$ search for the suitable parameters \{$\mathbf{y}' \leftarrow [k_0, k_1, z_1, z_2, \cdots, z_{L-1}]'\}$ that satisfy the objectives as described in Algorithm \ref{algMOTPE} as a set of non-dominated solutions in $O$ called the Pareto front. As a multiobjective problem, the non-dominated solution contains parameter choices that satisfy either or both objectives. In this paper, we selected the parameters set that minimizes both objectives. 

 \subsubsection{CADIC execution}

As the tuning of the CADIC parameters is centralized, the selected parameters \([k_0, k_1, z_1, z_2, \ldots, z_{L-1}]'\) are signaled to the subnetwork AP from a CADIC parameter tuning server after the tuning phase. These parameters are then used for the joint transmit power and sub-bands selection during the runtime of the subnetwork-controlled plants. At the start of the runtime phase, each subnetwork randomly selects an initial operating sub-band. This random selection ensures a balanced allocation of initial sub-bands across subnetworks. The subnetwork AP is tasked with executing the CADIC policy throughout the runtime to manage transmit power and select sub-bands at each transmission time interval. The subnetwork AP performs the following steps to execute CADIC during runtime:
\begin{enumerate}
    \item Transmit power selection: The AP estimates the stability condition or instantaneous control cost of its associated plant, denoted as $\chi$. Using $\chi$ and the tuned parameters $k_0$ and $k_1$, the AP applies the transmit power policy defined in \eqref{eqn:logistic} to determine the operating transmit power.
    \item Sub-band ranking: The AP continuously monitors the received signal strength across sub-bands. It ranks the sub-bands dynamically using the ranking function specified in \eqref{eqn:rank}.
    \item Sub-band Allocation: Based on the ranked sub-bands, the tuned parameters $z_1, z_2, \cdots, z_{L-1}$, and the estimate of $\chi$, the AP determines the operating sub-band(s) for the current transmission interval using \eqref{eqn:gL} and \eqref{eqn:subband_dec}.
    \item Schedule transmission: The AP schedules UL transmissions for the sensors within the subnetwork, and the DL transmission to the actuators. It communicates the transmission slots in the time-sub-band(s) grid and specifies the transmit power for the transmissions.
\end{enumerate}

In summary, the subnetwork AP autonomously executes the CADIC policy using locally available information, such as $\chi$, received signal strengths on the sub-bands, and the tuned parameter set $[k_0, k_1, z_1, z_2, \cdots, z_{L-1}]'$. This decentralized approach makes CADIC easy to implement within subnetwork-controlled plants.

It is important to distinguish between the CADIC parameters tuning phase and the execution phase. The parameters tuning phase is centralized as indicated in the objective \eqref{eqn:optiobj2} and described by Algorithm \ref{alg_CADIC}. It can be done offline at the central network using a well-calibrated simulator of the subnetwork-controlled plants or during an experimental start-up phase for the plants. During the tuning phase, the BO runs Algorithm 2, which returns the best values of the $k_0$,$k_1$ and $z_i$,$i=1,\cdots,L$ parameters. In the execution phase, these parameters are then used at each subnetwork independently, i.e. in a fully decentralized manner to select transmit power and operational sub-bands using equations \eqref{eqn:conP31} and \eqref{eqn:conP32}. 

\section{Simulation Settings}
\label{section_settings}
In this section, we will discuss the simulation assumptions of the subnetwork-controlled plants, including the channel propagation, radio, control plants model and the Bayesian optimization parameters. Table \ref{tab:systemparameters} summarises the simulation parameters. The simulation environment was implemented using the Python programming language.

\subsection{Deployment Settings}
\label{section_ass}

We consider a Monte Carlo simulation with several episodes, each lasting for $1000$ time steps, where each step is $1~\text{ms}$. At the start of each episode, $N$ subnetworks are randomly deployed in a factory area of $30 \times 30 ~\text{m}^2$.  Each subnetwork has a radius of $2~\text{m}$ and supports the closed-loop control of an associated plant (either Plant 1 or Plant 2) with dynamics as specified in (\ref{eqn:loop}). We assume the plant has a sensor (UL device) and an actuator (DL device) randomly positioned within the subnetwork area with the AP/controller at the centre. Next, we discuss the channel model, radio parameters, parameters of the control plants and the Bayesian optimization. 
\begin{table}[]
\begin{center}
  \caption{Simulation parameters}
\label{tab:systemparameters}
\begin{tabular}{m{4.2cm}m{3.5cm}} 
\toprule
\textbf{Parameter}                              & \textbf{Value}    \\
\toprule
\multicolumn{2}{c}{InF-S Deployment} \\ 
\toprule
Factory area, $L\times L$                              & 30~m x 30~m \\ 
Number of subnetworks, $N$       & 15        \\ 
Subnetwork radius, $R$                        & 2~m        \\ 
Num of devices per subnetwork, J     & 2         \\ 
Sensor to controller min distance  & 1~m        \\
Mobility model                    & RRWP \\
Velocity                           & 3~m/s \\
\toprule
\multicolumn{2}{c}{Channel model and radio parameters} \\ 
\toprule
SL clutter density,  $r$, clutter size, $ds$   & 0.35, 10~m       \\ 
Shadowing standard deviation,  LOS, NLOS & 4, 5.7       \\ 
Correlation distance, $dc$                       & 10~m       \\ 
Maximum transmit power, $P_{max}$                  & 0~dBm  \\ 
Center frequency, $f$                                 & 10~GHz      \\ 
Subcarrier spacing SCS                              & 480 KHz \\
Number of sub-bands                              & 3 \\
Sub-band size                                    & 3~channel blocks (36~SCs $\approx$ 17.3MHz) \\
Noise figure, NF                               & 10~dB      \\ 
Sensor data size                               & 64 bytes      \\ 
Control signal data size                               & 32 bytes      \\ 
Code rate                               & 1/2      \\ 
Frame length                                 & 1~ms \\
UL, DL latency                               & 0.1~ms, 0.1~ms      \\ 
Traffic                                      & Periodic \\
Interarrival Time                            &Plant 1 - 1~ms, Plant 2 - 3~ms \\
\toprule
\multicolumn{2}{c}{Control Plants Parameters} \\ 
\toprule
Plant 1: A, B                & \scriptsize
$ \begin{bmatrix}
0 & 1 & 0 & 0\\
0 & 0 & -1.4 & 0 \\
0 & 0 & 0 & 1 \\
0 & 0 & 168 & 0
\end{bmatrix}, \begin{bmatrix}
0\\
1.90\\
0\\
-28.57
\end{bmatrix}$          \\ 
Plant 2: A, B                  & \scriptsize
$ \begin{bmatrix}
0 & 1 & 0 & 0\\
0 & 0 & -1.4 & 0 \\
0 & 0 & 0 & 1 \\
0 & 0 & 84 & 0
\end{bmatrix}, \begin{bmatrix}
0\\
1.90\\
0\\
-14.29
\end{bmatrix}$      \\

Cost matrices Q, R: & \scriptsize
$  \begin{bmatrix}
1 & 0 & 0 & 0\\
0 & 10 & 0 & 0 \\
0 & 0 & 10 & 0 \\
0 & 0 & 0 & 100
\end{bmatrix}, \begin{bmatrix}
0.1\\
\end{bmatrix}$      \\ 
\end{tabular}  
\end{center}
\end{table}
\subsubsection{Channel model and radio parameters}
The In-factory subnetwork scenario is similar to an Industrial Internet of Things (IIoT) scenario where the base station (BS) and user terminal (UT) antennas are clutter-embedded as they are installed within the plants.
We adopt the 3GPP TR 38.901 IIoT channel model  \cite{Jiang2021,3GPP}, including the line-of-sight (LOS) and non-line-of-sight (NLOS) models for a sparse-clutter low-antenna (SL) scenario.  Sparse clutter characterizes an environment with large-sized clutters \cite{3GPP}. The link path-loss $PL$ is specified using the alpha-beta-gamma (ABG) model \cite{3GPP}. The LOS probability is calculated as a function of the link distance $d$, clutter size $ds$, and clutter density $r$ as in the 3GPP InF model \cite{3GPP}. The large-scale fading, $\rho$ in dB, is the sum of the link's $PL~\text{(dB)}$ and shadowing $\text{(dB)}$. We employ the spatially correlated shadowing model used in \cite{Adeogun,Lu2016}. The small-scale fading $h \sim \mathcal{CN}(0,1)$ is assumed to be Rayleigh-distributed. The time correlation of $h$ is described by Jake's Doppler model \cite{Jakes1994-je}, with the subnetworks moving randomly using a restricted random waypoint model (RRWM) at a speed of $3~\text{m/s}$.  We assumed that OFDM is used for the transmissions in the subnetwork and the UL and DL air latency is set to $0.1~\text{ms}$.  We consider a relatively large subcarrier spacing of $480~\text{KHz}$ similar to previous works on subnetworks \cite{Adeogun2020,Li2023} to support the sub-millisecond latency. The sensor data is set to $64~\text{bytes}$ which is within the range of low-end sensors in industrial applications \cite{Times6G,Rasmus2021}, and the control signal is set to $32~\text{bytes}$. We assume a code rate of $1/2$ and a frame length of $1~\text{ms}$ with the transmission error calculated as in \eqref{eqn:errormodel}. We consider $3$ sub-bands, each consisting of $3$ channel blocks.  We assume periodic traffic of inter-arrival time set to either $1~\text{ms}$ or $3~\text{ms}$ depending on whether the associated plant is an isochronous application or non-isochronous application \cite{Times6G}. The control cycle for isochronous applications is specified to be between $0.1~\text{ms}$ and $2~\text{ms}$ while non-isochronous applications require $3~\text{ms}$ to $20~\text{ms}$ cycle time \cite{Times6G}.
The choice of traffic type, data sizes and inter-arrival time are also aligned with the experimental IIoT data traffic analysis for unit test cells reported in \cite{Rasmus2021}.

\subsubsection{Control plants}
\begin{figure}[ht!]
    \centering
    \input{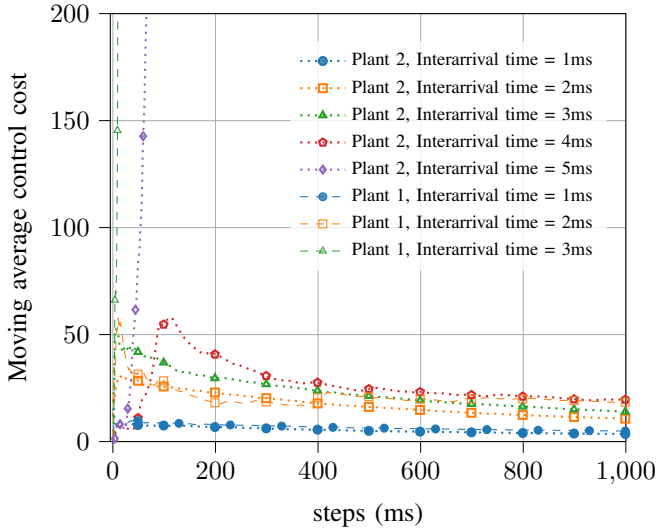}
    \caption{Analysis of the response of plant 1 and plant 2 to different inter-arrival times for perfect channel condition. Plant 2 requires a periodic closed-loop control at least every $4~\text{ms}$, while Plant 1 requires periodic closed-loop control at least every $2 ms$ to maintain a low control cost over a finite horizon.}
    \label{fig:interarrival}
\end{figure}

We consider $2$ different types of control plants with nonlinear dynamics \cite{Pedro2022}. A subnetwork may be associated with either of the plant types. The plants demand quick control cycles to maintain stability \cite{Eisen2019}. The response of the plants to periodic inter-arrival time is shown in Figure \ref{fig:interarrival}. Plant $1$ is an isochronous plant which requires periodic updates of less than $2~\text{ms}$ \cite{Times6G}. Plant $2$ is a non-isochronous plant which requires periodic updates of less than $4~\text{ms}$ \cite{Times6G}. The plants have four state variables, hence $\mathbf{x} \in \mathbbm{R}^4$ initialized from a uniform distribution $\mathcal{U}(-0.2,0.2)^4$. The action variable, $\mathbf{u} \in \mathbbm{R}^1$. The parameters of the control plants are given in Table \ref{tab:systemparameters}.

\subsubsection{Bayesian optimization configurations}
We implement the training Algorithm \ref{algMOTPE} using the Optuna optimization framework \cite{Optuna}, with $400$ trials, and $\mathcal{S} = 100$ startup trials, and $\mathcal{C} = 24$. The quantile $\gamma$ is automatically calculated and fine-tuned in the Optuna framework. The objectives are collected from a simulation of subnetwork control systems considering $E = 20$ episodes, finite horizon $1000$ time steps. We set $K_0 = [0,1]$ and $K_1 = (0,100]$, $Z_1 = (k_1, 200]$ and $Z_2 = (z_1, 300]$ for the parameter search space, where $[a,b]$ denotes a continuous closed interval including $a$ and $b$, while, $(a,b]$ denotes a continuous half-closed interval not including $a$ but including $b$.

\subsection{Benchmarks}
\label{section_bench}
We compared our proposed scheme, CADIC to the following benchmarks:
\begin{itemize}

\item Ideal Scenario (Ideal) - In the ideal scenario, we assume that the BLER is always $0$, which implies that communication is always successful. This shows the upper bound for the control performance given the system dynamics, the random noise in the sensor observation and the cost matrix. 

\item SISA - A centralized sub-band selection sequential algorithm proposed in 
 \cite{Li2023}. It is an iterative algorithm that minimizes the sum of the interference-to-signal power ratio. For its operation, each subnetwork must measure the channel gain of the interference link from all other subnetworks over all sub-bands, and report to a central entity that computes SISA. The central entity reports the allocated sub-band back to the subnetworks. We consider periodic allocation at every $10$ time steps which we numerically confirmed to be sufficient. Note that the channel gain information is collected from the UL phase.
 
\item SISA with Power Control (SISA+PC) - This is a centralized method that consists of two steps. In the first step, the subnetworks are allocated sub-band using SISA. In the second step, transmit power control is performed for the subnetworks in the same sub-band to maximize a fair single objective function of rate for all subnetworks using sequential least square quadratic programming. For this, we consider the product of the rate as the fairness metric. Like SISA, this algorithm is performed every $10$ time steps.

\item Sequential Greedy Distributed Sub-band Selection (Seq Greedy) - This is a distributed sub-band selection algorithm, proposed in \cite{Bagheri2024}. Each subnetwork selects the sub-band with the lowest aggregated interference according to a predefined order. Subnetworks that have already selected a sub-band transmit a reference signal on that sub-band, which is then measured by the next subnetwork in line to make its selection. To adapt to dynamic channel conditions caused by the random mobility of the subnetworks, we perform periodic execution of the algorithm every $10$ time steps.

\item Random Sub-band Allocation (Random) - Each subnetwork randomly selects a sub-band at every time step with a fixed maximum transmit power.

\item Fixed power, All sub-bands (FP) - The subnetworks operate over all the sub-bands using a maximum transmit power set to $0~dBm$.
\end{itemize}

\section{Main Results And Discussion}
\label{section_result}
 In this section, we first discuss how CADIC coordinates inter-subnetwork interference. Afterwards, we compare the performance of the algorithm to the benchmark algorithms in terms of the finite horizon control cost, and the BLER. Finally, we analyze their scalability to different subnetwork densities in terms of the consistency in control performance, the execution complexity, and the sensitivity of the trained CADIC policy. To analyze the performance of the algorithms, we generally consider a Monte Carlo simulation of $9000$ episodes and $N=15$ subnetworks. 

\begin{figure}[htbp]
    \centering
    \subfloat[]{\input{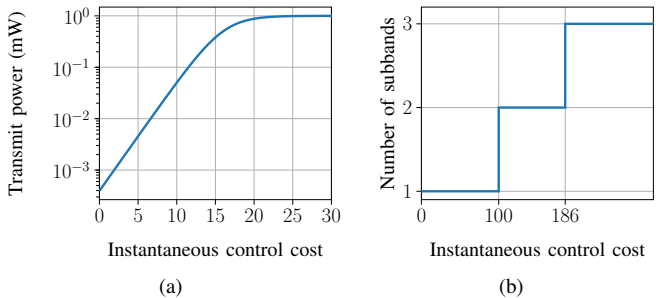}}
    \hspace{5pt}
    \subfloat[]{\begin{tikzpicture}[scale=0.45]

\definecolor{darkgray176}{RGB}{176,176,176}
\definecolor{steelblue31119180}{RGB}{31,119,180}

\begin{axis}[
tick align=outside,
tick pos=left,
x grid style={darkgray176},
xmajorgrids,
xmin=0, xmax=300,
xlabel={Instantaneous control cost},
xtick style={color=black},
xtick={0, 100, 186},
y grid style={darkgray176},
ymajorgrids,
ymin=0.9, ymax=3.2,
ylabel = {Number of subbands},
ytick style={color=black},
ytick={1, 2, 3},
label style={font=\LARGE},
tick label style={font=\LARGE},
xlabel style={yshift=-1.0em}  
]
\addplot [thick, line width=2pt, steelblue31119180]
table {%
0 1
100 1
100 2
186 2 
186 3
200 3
210 3
220 3
230 3
240 3
250 3
260 3
270 3
280 3
290 3
300 3
310 3
320 3
330 3
340 3
350 3
360 3
370 3
380 3
390 3
400 3
410 3
420 3
430 3
440 3
450 3
460 3
470 3
480 3
490 3
};
\end{axis}

\end{tikzpicture}}
    \caption{CADIC Policy for transmit power control and sub-band allocation. (a) Transmit power as a logistic function of instantaneous control cost, and parameters $k_0=0.49$ and $k_1=16$. (b) Number of sub-bands to be selected as a piecewise step function of the instantaneous control cost with parameters  $z_1=100$ and $z_2=186$.}
    \label{fig:configed_param}
\end{figure}

\begin{figure}[htbp]
    \centering
    \includegraphics[width=0.45\textwidth]{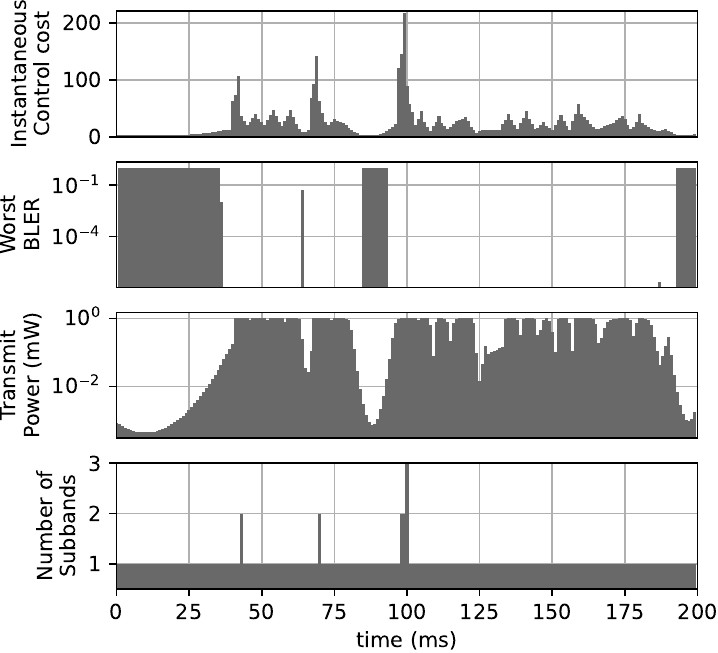}
    \caption{Explanation of CADIC operation for a randomly selected plant wirelessly controlled by subnetwork. CADIC reduces the transmit power and number of operating sub-bands when the plant is more stable (low instantaneous control cost), and increases the transmit power and/or the number of sub-bands when the plant is less stable.}
    \label{fig:cadic_explained}
\end{figure}

\begin{figure*}[htbp]
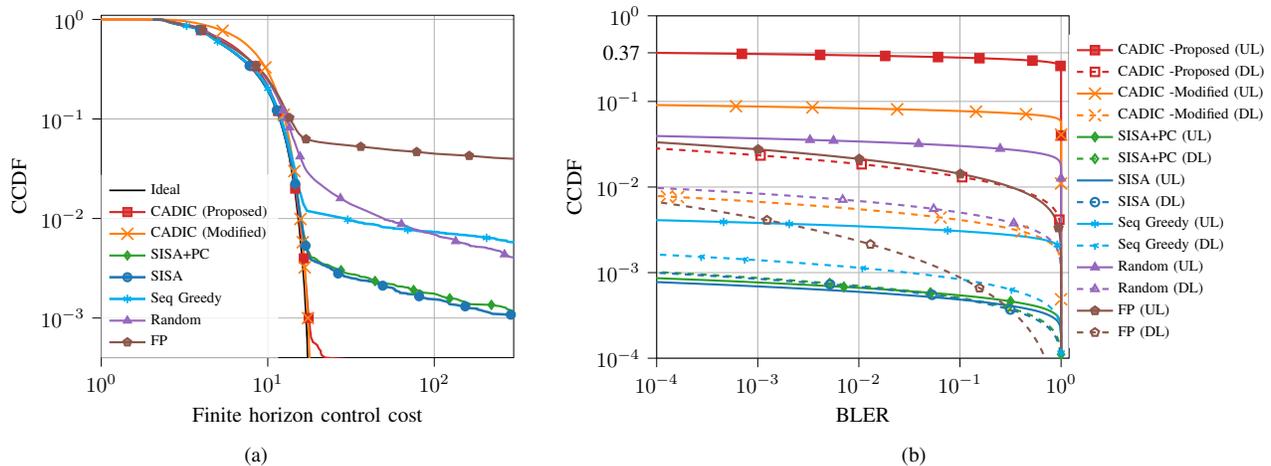

    \centering
    \subfloat[]{\input{results/mean_lqr.tex}}
    \hspace{10pt}
    \subfloat[]{\input{results/BLER.tex}}
    \caption{Control metric performance vs radio-metric performance for $N=15$ subnetworks (a) CCDF of the finite horizon control cost. (b) CCDF of the BLER.}
    \label{fig:lqr_bler}
\end{figure*}

\subsection{Description of Interference Coordination using CADIC}
\label{section_example}
Tuning the parameters of CADIC using Algorithm \ref{alg_CADIC} considering the default simulation settings resulted in the values $k_0=0.49$, $k_1=16$, $z_1=100$ and $z_2=186$ which minimizes both objectives in \eqref{eqn:optiobj3}. Figure \ref{fig:configed_param} (a) and (b) show the transmit power and number of sub-bands as a logistic function and a step function of the plant's instantaneous control costs respectively. As illustrated in Figure \ref{fig:configed_param} (a), the subnetwork associated with the plant with a low instantaneous control cost generally operates on low transmit power, the transmit power increases for a higher instantaneous control cost, reaching the maximum transmit power around an instantaneous control cost of $20$. Similarly, Figure \ref{fig:configed_param} (b) shows that subnetworks associated with plants with low instantaneous control costs operate on only a $1$ sub-band, however, the number of sub-bands to be used increases step-wise as the instantaneous control cost increases. 

To analyze the operations of CADIC, we consider an example of a subnetwork-controlled plant in Figure \ref{fig:cadic_explained} showing the first $200 ~\text{ms}$. As evident in the figure, the plant starts from a stable state and can operate on an open loop for almost $30 ~\text{ms}$. In the stable condition i.e. low instantaneous control cost, the BLER is high, because the CADIC algorithm assigned low transmit power to the associated subnetwork in this regime, hence the plant causes negligible interference to its neighbouring plants, which may be unstable and can benefit from the low interference. On the other hand, when the instantaneous control cost increases, the transmit power in the subnetwork increases simultaneously, and to cope with worsening control cost, the subnetwork selects more sub-bands momentarily to achieve an extremely low BLER guaranteeing the successful closed-loop control of the associated plant back to stable regime.  In general, we can observe that multiple sub-bands are selected seldom only when the plant is in a critical condition,  and then the subnetwork switches back to one sub-band with high transmit power as the plant recovers. Also, the transmit power is more sensitive to granular changes in the instantaneous control cost in the lower control cost regime. Consequently, in the long term, the finite horizon control cost is minimized. The operation of CADIC inherently promotes fairness in a distributed systems of subnetwork-controlled plants. It dynamically adjusts the allocation of radio resources by reducing the share allocated to subnetwork-controlled plants that have already achieved stability. This reduction minimizes interference and, in turn, increases the resources available to subnetwork-controlled plants that are unstable. In this way, it ensures a goal-oriented and fair utilization of the limited radio resources, balancing system stability and efficient reuse of the radio resource.

\subsection{Performance Comparison for Finite Horizon Control Cost and BLER}
\label{section_controlvsradio}
Figure \ref{fig:lqr_bler} shows the Complementary Cumulative Distribution Functions (CCDF) of the finite horizon control cost and the BLER. Our proposed CADIC algorithm demonstrates performance comparable to the Ideal case up to the 99.9th percentile, as illustrated in Figure \ref{fig:lqr_bler}(a). Moreover, CADIC consistently outperforms the decentralized sub-band selection scheme (Seq Greedy) and the centralized schemes (SISA and SISA+PC), which are optimized specifically to minimize the BLER. However, the benefits of optimizing for BLER are evident when comparing Seq Greedy, SISA, and SISA+PC to the un-optimized approaches of Random sub-band allocation scheme and FP. Seq Greedy, SISA, and SISA+PC generally achieve better BLER performance and reduced control costs than Random and FP. Furthermore, the centralized schemes (SISA and SISA+PC) outperform the decentralized Seq Greedy approach, not only in minimizing BLER but also in achieving superior finite-horizon control cost performance.

However, a notable conflict between the radio-metric BLER and the controlled plants' stability is observed when we compare SISA, and SISA+PC to CADIC. While CADIC outperforms SISA in the control metric, it generally performs worse than SISA and SISA+PC in BLER metrics. However, this conflict can be demystified considering that; (1) The BLER statistics are collected at every transmission instance and generally favour algorithms like SISA that minimize the BLER to ensure the periodic successful transmission of the sensor observation and the control signal. (2) As seen in Figure \ref{fig:cadic_explained}, CADIC generally allocate low radio resources (transmit power and sub-bands) to subnetworks associated with stable plants, leading to high BLER in the stable regimes. We can decide that subnetworks allocated low transmit power should rather not transmit. Given this, we devise CADIC (Modified), where subnetworks allocated transmit power less than $-25\text{dBm}$ are forced to not transmit, hence this further reduces interference. As seen in the Control cost performance CADIC (Modified) achieves a better performance than CADIC (Proposed) in the lower CCDF percentiles, with a slight loss in the upper percentile of the CCDF curve, and it equally achieves better BLER than CADIC (Proposed).

This result challenges the assumption that the RRM algorithm achieving the best BLER curve is inherently the most effective way to ensure quality of service for wireless applications, such as the wireless closed-loop control of plants. While minimizing BLER can enhance periodic successful transmissions, it often leads to overprovisioning of radio resources, thereby limiting the number of subnetwork-controlled plants that can be reliably supported with finite radio resources. In contrast, the goal-oriented approach of CADIC focuses on making more radio resources available and ensures reliable communication for goal-relevant transmissions. By dynamically reducing radio resource allocation for stable subnetwork-controlled plants and reallocating those resources to subnetwork-controlled plants on the verge of instability, CADIC enables more efficient spatial reuse of limited radio resources, as demonstrated by its ability to achieve lower control costs with high reliability. 

The idea of the CADIC (Proposed) or CADIC (modified) also helps to save transmit power compared to the other schemes that strive to ensure successful periodic transmission. For example, in the case of CADIC (Modified), we numerically identified that transmissions only occur $60\%$ of the time compared to SISA. 

It is also important to note why the BLER DL is generally equal to or lower than UL. This is because the control signal data for the DL is 2$\times$ smaller than the sensor data for UL. In addition, the SINR for DL is either close or much better than the UL because of the following reasons. (1) Subnetworks are short in range, both the UL and DL channels experience relatively the same desired link channel gain. (2) DL interference may be less since the DL transmission is only scheduled if the controller receives the UL packet successfully, hence only a subset of the subnetworks that transmitted UL may interfere in the DL phase. Hence, the transmit power and sub-band allocation decision using the information of the desired link and interference links of the UL device is sufficient for the DL device in the subnetwork.





\begin{figure}[t]
    \centering
    \includegraphics[width=0.46\textwidth]{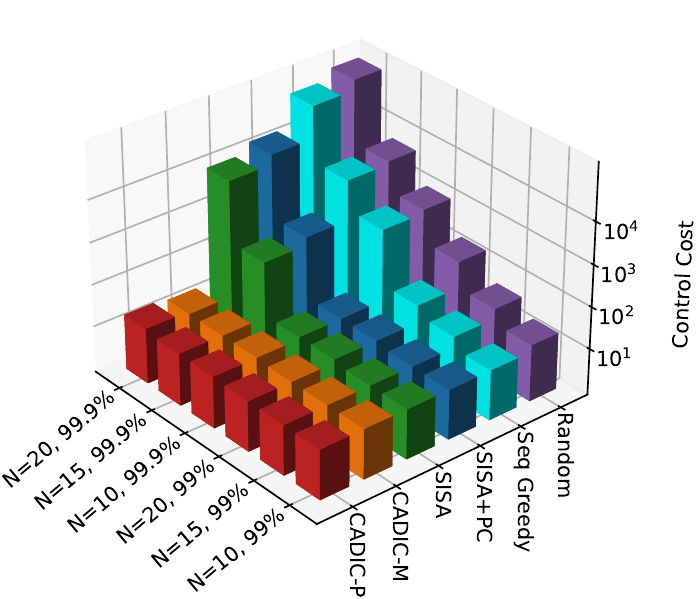}
    \caption{99\%, 99.9\% Finite horizon control cost for different deployment density $N=10, 15, 20$ achieved by the different algorithms (Note CADIC-P is the same as CADIC (Proposed) and CADIC-M is the same as CADIC (Modified).}
    \label{fig:density_control_cost}
\end{figure}

\subsection{Scalability Analysis}
\label{section:scalability}
We consider scalability analysis for the different algorithms by analyzing finite-horizon control cost for different densities of subnetworks deployment. We also analyzed the execution complexity and the sensitivity of the trained parameters in the case of CADIC to the various changes in density.

\subsubsection{Control performance for different subnetwork density}
\label{section_compare_diff_density}
One of the important criteria for an inter-subnetwork interference management algorithm is how well it can maintain a good performance when the number of subnetworks changes. In Figure \ref{fig:density_control_cost}, we compare the $99.9\%$ and $99\%$ control cost for $N=10$, $N=15$ and $N=20$ subnetworks in the fixed factory area. As seen in the Figure, CADIC uses the same parameters from the training conducted in settings of $15$ subnetworks, yet it generally maintains a reliably low control cost similar to the ideal scenario for all the densities including the $99.9\%$ percentile up to $N=20$ subnetworks. However, SISA and SISA-PC already failed to maintain a low control cost, especially in the higher percentile of $99.9\%$ for $N=15$ and $N=20$. Similarly, Seq Greedy also failed to maintain a low control cost at higher percentile as the density of the subnetwork deployment increases.
This result highlights the efficiency of CADIC in increasing the number of subnetwork-controlled plants that can be sustained with the same amount of radio resources compared to SISA and SISA-PC. In Table \ref{tab:sensitivity}, we examine the scenario where the subnetwork density differs between the training and testing settings. It is important to analyze how sensitive the trained parameters of CADIC are to such changes. In addition to our initial setup of $15$ subnetworks, we consider a less dense scenario of $10$ subnetworks and a more dense scenario of $20$ subnetworks, both in the training and execution phases. The $99.9\%$ finite-horizon control cost results of the trained parameters for the different densities are shown in Table \ref{tab:sensitivity}. For the analysis, we consider parameters that minimize the mean finite-horizon control costs across all plants. While there appear some changes in the trained parameters for the different training configurations, it is interesting to see that the $99.9\%$ performance is relatively the same even though the training and execution densities are different.

In addition, we analyze the balance in the distribution of operating sub-band selection. In a highly dense scenario of $20$ subnetworks, we observed a relatively balanced sub-band selection. The difference between the most and least used sub-bands was less than $3\%$, indicating that the CADIC approach maintains statistical balance in sub-band selection even in very dense deployment.

\renewcommand{\arraystretch}{1.0}
\begin{table}[]
\caption{99.9\% finite-horizon control cost for different densities of subnetworks in training and execution phases, and the trained parameters $k_0, k_1, z_1, z_2$.}
\resizebox{\columnwidth}{!}{%
\begin{tabular}{lcccc}
 &
  \multicolumn{4}{c}{Execution N} \\
 &
  \cellcolor[HTML]{ECF4FF} &
  \cellcolor[HTML]{ECF4FF}10 &
  \cellcolor[HTML]{ECF4FF}15 &
  \cellcolor[HTML]{ECF4FF}20 \\ \cline{3-5} 
 &
  \multicolumn{1}{c|}{\cellcolor[HTML]{DAE8FC}Parameters} &
  \multicolumn{1}{c|}{\cellcolor[HTML]{DAE8FC}\begin{tabular}[c]{@{}c@{}}0.4, 20,\\ 96, 146\end{tabular}} &
  \multicolumn{1}{c|}{\cellcolor[HTML]{DAE8FC}\begin{tabular}[c]{@{}c@{}}0.49, 16,\\ 100, 186\end{tabular}} &
  \multicolumn{1}{c|}{\cellcolor[HTML]{DAE8FC}\begin{tabular}[c]{@{}c@{}}0.66, 14,\\ 130, 246\end{tabular}} \\ \cline{3-5} 
\multicolumn{1}{c}{} &
  \multicolumn{1}{c|}{\cellcolor[HTML]{ECF4FF}10} &
  \multicolumn{1}{c|}{17.00} &
  \multicolumn{1}{c|}{17.60} &
  \multicolumn{1}{c|}{18.10} \\ \cline{3-5} 
\multicolumn{1}{c}{} &
  \multicolumn{1}{c|}{\cellcolor[HTML]{ECF4FF}15} &
  \multicolumn{1}{c|}{17.04} &
  \multicolumn{1}{c|}{17.64} &
  \multicolumn{1}{c|}{20.74} \\ \cline{3-5} 
\multicolumn{1}{c}{\multirow{-3}{*}{Train N}} &
  \multicolumn{1}{c|}{\cellcolor[HTML]{ECF4FF}20} &
  \multicolumn{1}{c|}{17.60} &
  \multicolumn{1}{c|}{17.00} &
  \multicolumn{1}{c|}{18.00} \\ \cline{3-5} 
\end{tabular}%
}
\label{tab:sensitivity}
\end{table}

\subsubsection{Execution Complexity Analysis}
\label{section_execution_complexity}
The execution complexity of the interference coordination algorithm can be analyzed in terms of the number of radio-sensing operations required by the subnetwork to obtain the information needed by the interference coordination algorithm, the number of radio messages that must be exchanged between the subnetwork and a centralized interference manager in the case of the centralized schemes and finally the asymptotic runtime complexity in executing the algorithm. While Random and FP may not incur any cost, it is expected that the optimized algorithm including CADIC, Seq Greedy, SISA, and SISA+PC would incur some execution complexity.

In terms of the number of radio-sensing operations, CADIC only requires sensing the cumulative interference on $L$ sub-bands, hence the cost of $L$ radio-sensing operations. This approach is similar to Seq Greedy, where the subnetwork must sense the cumulative interference across the $L$ sub-bands. However, since Seq Greedy operates in a predefined round-robin order, each subnetwork must wait for its turn, resulting in a radio-sensing cost that grows linearly with $N$. On the other hand, the centralized approach of SISA and SISA+PC require that each of the subnetworks senses the interference on each mutual interfering link and for each sub-band, hence each subnetwork would incur a cost $L(N-1)$ number of radio sensing operations to measure the interference on the mutual interfering channel from its $N-1$ neighbours for the $L$ sub-bands, accumulating a cost of $NL(N-1)$ for all subnetworks. Hence, SISA and SISA+PC incur a radio-sensing cost that grows quadratically with the number of subnetworks, while Seq Greedy experiences a linear growth in radio-sensing cost. In contrast, CADIC achieves the least radio-sensing cost, independent of the number of subnetworks.

For SISA or SISA+PC, the subnetworks must also signal the measured interference on all sub-bands from all neighbours, including the channel gain on their desired links to the central entity. The total number of signalling messages is $LN^2$. Also, the central entity would be required to signal the sub-band decision back to the subnetworks which will cost $N$ signalling messages, hence a total of $LN^2 + N$ signalling messages. Hence, SISA would incur signalling costs that grow quadratically as the number of subnetworks. In Seq Greedy, each subnetwork must transmit a reference signal on its selected sub-band until the last subnetwork in the sequential order finalizes its sub-band selection, resulting in a total of $N$ signaling messages. On the other hand, CADIC is decentralized at the subnetwork level, hence it incurs no signalling cost for execution. 

For the asymptotic computation complexity, CADIC being decentralized only incurs $O(1)$ computation cost, similar to Seq Greedy concerning the growing number of subnetworks. On the other hand, SISA is an iterative algorithm and it is said to converge in about $10$ iterations \cite{Li2023}. One iteration requires operations in an inner loop of the number of subnetworks, hence the number of computations in this inner loop grows linearly as the number of subnetworks, hence an asymptotic computation cost of $O(N)$.

This execution analysis sheds more light on the low complexity of CADIC compared to SISA, SISA+PC and Seq Greedy in terms of execution, further elaborating on the efficacy of CADIC for managing inter-subnetwork interference in large-scale deployment of subnetwork-controlled systems.

\section{Future Work}
\label{sect_futurework}
This study focused on subnetworks supporting closed-loop control applications; however, in-factory subnetworks may serve diverse applications, such as data gathering and collaborative system. Future research could explore developing goal-oriented RRM approaches tailored to these varied applications, addressing their unique requirements, dynamics and constraints. While we applied BO to tune the parameters of CADIC in this study, BO can become computationally expensive as the number of tunable parameters increases, such as in scenarios with a larger number of sub-bands. To address this limitation, future studies could explore the use of online learning algorithms, such as reinforcement learning, as an alternative to offline BO for parameter tuning. Currently, we assume that sub-packets are uniformly distributed across the operating sub-bands and that transmit power is evenly divided among them. Future work could investigate more flexible sub-packet distribution and power allocation strategies, where different operating sub-bands are assigned varying transmit power levels to optimize performance under diverse interference conditions. Furthermore, a comprehensive theoretical analysis of the trade-offs between code length, interference, and error probability is needed. Such analysis could also include the development of closed-form models to quantify the impact of inter-subnetwork interference, available radio resources, and subnetwork-controlled plant performance, deriving theoretical limits and optimal RRM configurations. To further enhance the subnetworks' capacity for supporting closed-loop control applications with higher reliability, a hybrid approach could be explored. This would involve combining the decentralized CADIC algorithm with intermittent centralized interference coordination mechanisms. Additionally, beyond optimizing communication parameters alone, future research could investigate joint optimization of both communication and control parameters. For example, optimizing control parameters such as the sampling interval alongside radio resource selection.

\section{Conclusion}
\label{section_conclusion}
In this paper, we investigated the inter-subnetwork interference mitigation problem in dense subnetworks deployed for closed-loop control of plants in the factory. We propose a decentralized goal-oriented joint power and sub-band allocation policy CADIC that exploits the information of the states of the wireless controlled plants to manage interference between the subnetworks and minimize the long-term control cost. We proposed a Bayesian optimization procedure based on a tree-structured Parzen estimator to learn the parameters of the policy. We showed that CADIC can support a factor of $\times2$ the density of subnetwork-controlled plants with high reliability compared to the conventional inter-cell interference management approach that aims to minimize the BLER. In addition, CADIC has low complexity and can be executed quickly to cope with the changes in channel conditions due to the mobility of the subnetworks and the changing radio resource needs of the subnetwork-controlled plants. As wireless concepts begin to dive into more challenging applications like wireless control of manufacturing plants, and with the advent of application-specific short-range cells such as subnetworks, this paper highlights the importance of exploring the knowledge of the underlying control application in managing interference between wireless cells to meet application goals with high reliability. Directions for future research were also identified.

\bibliographystyle{IEEEtran}
\bibliography{references}

\end{document}